\documentclass[10pt,journal,compsoc]{IEEEtran}

\ifCLASSOPTIONcompsoc
  \usepackage[nocompress]{cite}
\else
  \usepackage{cite}
\fi

\usepackage[dvips]{graphicx}
\usepackage{amsmath,algorithm}
\usepackage{amsthm}

\usepackage[noend]{algpseudocode}
\usepackage{setspace}
\usepackage{color}
\usepackage{graphics}
\usepackage{tikz}
\usetikzlibrary{arrows}
\usepackage{cases}
\usepackage{relsize}
\usepackage{tablefootnote}
\usepackage{threeparttable}
\usepackage{booktabs}
\usepackage{makecell}
\usepackage[markup=hmarkupi]{changes}

\usetikzlibrary{arrows}

\makeatletter
\def\endthebibliography{%
  \def\@noitemerr{\@latex@warning{Empty `thebibliography' environment}}%
  \endlist
}
\makeatother

\DeclareMathOperator{\Exp}{Exp}

\newcommand{\T}[1]{\textrm{#1}}

\newcommand{\M}[1]{{\bf #1}}

\newcommand{\bl}[1]{{\textcolor{black}{ #1}}}

\begin{document}

\title{Diversity of structural controllability of complex networks with
given degree sequence}

\author{Abdorasoul~Ghasemi,
        M\'arton~P\'osfai, and~Raissa M. D'Souza
\IEEEcompsocitemizethanks{\IEEEcompsocthanksitem A. Ghasemi is with the Department
of Computer Engineering, K. N. Toosi University of Technology, Tehran, Iran, E-mail: arghasemi@kntu.ac.ir.

\IEEEcompsocthanksitem M\'arton~P\'osfai  is with Complexity Sciences Center and Department of Computer Science, University of California, Davis, CA 95616, USA.
\IEEEcompsocthanksitem Raissa M. D'Souza is with Complexity Sciences Center, Department of Computer Science and Department of Mechanical and Aerospace Engineering, University of California, Davis, CA 95616, USA and Santa Fe Institute, 1399 Hyde Park Road, Santa Fe, NM 87501,  USA.
}

\thanks{Manuscript received July xx, 2019; revised July xx, 2019.}}

\markboth{IEEE Transactions on Network Science and Engineering}%
{Shell \MakeLowercase{\textit{et al.}}: Bare Demo of IEEEtran.cls for Computer Society Journals}

\IEEEtitleabstractindextext{%
\begin{abstract}
We investigate to what extent the degree sequence of a directed network constrains the number of driver nodes.
We develop a pair of algorithms that take a directed degree sequence as input and aim to output a network with the maximum or minimum number of driver nodes. 
We find an upper bound for the maximum and a lower bound for the minimum, and show that the algorithms achieve these bounds for all real and model networks, with few exceptions characterized by tiny system size and heterogeneous degree distributions.
Applying these algorithms to a broad range of real networks, we show the gap between the upper and lower bounds can vary dramatically across different degree sequences. Thus, we introduce the notion of \bl{structural control} complexity to capture how much more difficult it is to control a specific network beyond what is required given its degree sequence, suggesting additional structure is present.
Using model networks, we numerically and analytically investigate how typical features of the degree distribution affect the range of required driver nodes. We find that the minimum is determined by the number of sources or sinks, while the maximum is strongly affected by the presence of hubs.
\end{abstract}

}

\maketitle

\IEEEdisplaynontitleabstractindextext

\IEEEpeerreviewmaketitle

\IEEEraisesectionheading{\section{Introduction}\label{sec:introduction}}

The interface of network science and control theory provides a means to understand underlying principles of controlling complex systems~\cite{liu2016control}. Social, biological, and human-made complex systems are composed of many interacting parts, and the structure of the networks formed by these interactions strongly influences their function, behavior, and resilience. Therefore, it is particularly interesting to seek out existing methods and to develop new methods of control theory that leverage the underlying network structure of dynamical systems~\cite{mesbahi2010graph, liu2011controllability, cornelius2013realistic, fiedler2013dynamics, yuan2013exact}. Such methods allow us to design strategies to influence the behavior of complex systems and to characterize underlying mechanisms that inhibit or enhance control.

In this article, we investigate to what extent the degree sequence of a directed network constrains its controllability. Specifically, we develop algorithms to identify the maximum and minimum number of external signals necessary to control networks with a given degree sequence.
Leveraging these algorithms, we introduce the notion of \bl{structural control} complexity, a measure of network controllability that captures the role of additional level of structure, beyond just the degree sequence, that necessitates a network to need more control signals beyond what is dictated by the degree distribution alone. We then use these tools to systematically analyze a collection of real and model networks.

We rely on the framework of structural controllability of linear systems~\cite{lin1974structural}, which exploits the deep connection between graph combinatorics and linear algebra, allowing us to effectively study some control properties of directed networks. Specifically, we assume that a directed complex network with $N$ nodes is governed by linear time-invariant dynamics
\begin{equation}
\dot x(t) = \M A x(t) + \M B u(t),
\end{equation} 
where $x(t)\in R^N$ represents the state of the nodes, $\M A\in R^{N\times N}$ is the weighted adjacency matrix, $u(t)\in R^M$ represents $M$ independent control signals, and $\M B\in R^{N\times M}$ is the matrix that identifies how the control signals are coupled to the network.

A dynamical system is controllable if it can be driven in finite time from any initial state $x_0$ to any final state $x_1$ with a suitable choice of inputs $u(t)$. Traditional methods to determine controllability of a linear systems are impractical for large complex networks, because they require accurate knowledge of all link weights and are numerically unstable~\cite{liu2011controllability, sun2013controllability}. To overcome these difficulties, we turn to the theory of structural controllability. We say that matrix $\M A$ has the same structure as matrix $\M A^*$ if $\M A$ has zero elements wherever  $\M A^*$ does while its non-zero elements can have different values.

A linear system $(\M A^*, \M B^*)$ is structurally controllable if there exists a pair of matrices $\M A$ and $\M B$ with, respectively, the same structure as $\M A^*$ and $\M B^*$ such that $(\M A, \M B)$ is controllable. Importantly, if a network is structurally controllable, it is controllable for almost all link weight combinations~\cite{lin1974structural}. Therefore studying structural controllability of typical weighted directed networks is equivalent to studying controllability in the original sense. Although structural control theory was developed for simple directed networks~\cite{lin1974structural, liu2011controllability, gao2014target, iudice2015structural}, it has been extended to multiplex networks~\cite{menichetti2016control, posfai2016controllability}, temporal networks~\cite{pan2014structural, posfai2014structural}, link dynamics~\cite{nepusz2012controlling}, and most recently undirected networks~\cite{menara2018structural,mousavi2018structural}.

The underlying network structure of a complex system specifies $\M A$, while typically many choices of $\M B$ allow full control.
Often the minimum number of signals necessary for control is used to quantify the controllability of a network, i.e., the minimum $M$ such that there exists a $\M B \in R^{N\times M}$ rendering $(\M A, \M B)$ controllable~\cite{liu2011controllability, yuan2013exact, nepusz2012controlling}. Reference~\cite{liu2011controllability} introduced the concept of driver nodes as a minimal set of nodes that have to be controlled by independent signals directly to ensure controllability of the network. The minimum number of independent signals and driver nodes are equal; therefore, the two are used interchangeably in recent literature. Following convention, we denote the minimum number of independent signals or drivers as $N_\T D$.
 
Liu et al. mapped the problem of identifying $N_\T D$ of a directed network to finding the maximum matching in its bipartite representation~\cite{liu2011controllability}. Consider a directed network $G~=~(V,E)$, where $V$ is the set of $N$ nodes and $E$ is the set of $L$ directed links connecting these nodes. To construct its bipartite representation $\hat G=(V^+, V^-,\hat E)$, we split each node $v_i\in V$ into two copies $v_i^+\in V^+$ and $v_i^-\in V^-$ and we add an undirected link $(v_i^+-v_j^-)\in\hat E$ if there exists a directed link $(v_i\rightarrow v_j)\in E$ in the original network. A maximum matching $\hat E_\T{MM}\subset \hat E$ is a maximum cardinality set of links that do not share endpoints. The number of driver nodes is determined by
\begin{equation}
N_\T D = \max(N-\lvert\hat E_\T{MM}\rvert, 1).
\end{equation}
This mapping provides computationally efficient and numerically robust tools to study controllability of large complex networks. See Fig.~1a-b for an example.

The effect of typical structural properties of complex networks on $N_\T D$ has been thoroughly investigated. Using a collection of real and model networks, Liu et al. showed that the degree sequence of networks can largely predict $N_\T D$, and that degree heterogeneity inhibits control~\cite{liu2011controllability, menichetti2014network}.
The degree sequence of networks, however, does not completely determine their controllability. 
For example, Ref.~\cite{posfai2013effect} showed that beyond degree distribution, degree correlations of connected node pairs also affect $N_\T D$, while community structure and short-range loops added via randomized link-rewiring have little effect on controllability.
In addition, Wang et al. found that strategic addition of links to a network can drastically reduce $N_\T D$, while having only little effect on the overall degree distribution, further demonstrating that the degree sequence does not uniquely determine $N_\T D$ of a network~\cite{wang2012optimizing}.

Here, we ask a complementary question: Instead of investigating how structural properties affect $N_\T D$, we are interested in to what extent the degree sequence of a network constrains the maximum and minimum value of $N_\T D$. In Sec.~\ref{sec:maxminprob}, we introduce a pair of algorithms that take a directed degree sequence as input and aim output $ G_\T{max}$ and $G_\T{min}$, a pair of networks with the maximum and minimum number of driver nodes with that degree sequence.
We show that these algorithms output demonstrably correct results for realistic model networks and a diverse collection of real networks. Identifying the maximum and minimum number of driver nodes allows us to introduce the notion of \bl{structural control} complexity, a measure of controllability that takes into account the constraints of the degree sequence of networks.

In Secs.~\ref{sec:realnets} and \ref{sec:model_nets}, we apply our algorithms to systematically investigate the possible range of $N_\T D$ and the \bl{structural control} complexity of real and model networks.
In Sec.~\ref{sec:prev_work}, we discuss the relationship between our results and previous work. Specifically, we probe the possible structure of $G_\T{max}$ and $G_\T{min}$ by adding degree correlations through link rewiring, and we also apply our results to understand how the degree sequence of a network constrains the control profile of the network~\cite{ruths2014control, campbell2015topological}. In this paper, we provide a range of findings that extend and complement our current knowledge on the relation of network structure and controllability, providing new insights and allowing deeper understanding of previously established results. 

\section{Maximum and minimum driver nodes}\label{sec:maxminprob}

In this section, we introduce the problem of constructing networks with maximum and minimum number of driver nodes and we develop algorithms to solve them. 
First, consider a bi-degree sequence (BDS), i.e., $N$ pairs of integers $(k^+_1,k^-_1), \ldots, (k^+_N,k^-_N)$, where $k^+_i$ and $k^-_i$ are  the assigned out- and in-degree of node $v_i$, respectively. A BDS is graphical, if there exists a directed network $G=(V,E)$ with the given degree sequence, such that it does not contain double links, while self-loops are allowed. Network $G$ is called a graphical realization of the BDS.

To determine if a BDS is graphical or not, we use the Havel-Hakimi (HH) algorithm~\cite{havel1955remark,hakimi1962realizability}. In addition, if the BDS is graphical, the HH algorithm constructs the bipartite representation $\hat G = (V^+,V^-,\hat E)$ of a graphical realization. 
We start with two sets of $N$ unconnected nodes, $V^+$ and $V^-$. 
We assign $k^+_i$ out-stubs to each node $v_i^+\in V^+$ and $k^-_i$ in-stubs to each node $v_i^-\in V^-$. 
We now pick a node $v_i^+ \in V^+$ and we reduce $V^-$ by $v_i^+$, that is we form links by connecting the out-stubs of $v_i^+$ to the $k^+_i$ nodes in $V^-$ that have the most unused in-stubs. If there are less than $k^+_i$ available nodes in $V^-$, the process is unsuccessful. We repeat this step for all nodes in $V^+$.
Generally, we say $S \subseteq V^-$  is reducible by $R \subseteq V^+$ if we can reduce $S$ by all nodes in $R$ iteratively. The BDS is graphical if we can reduce $V^-$ by the nodes in $V^+$ and there are no unconnected stubs in $V^-$ remaining.

The HH algorithm creates one graphical realization; generally, however, there are many realizations of a BDS. We are interested in finding a realization $G_\T{max}$ ($G_\T{min}$) that requires the most (least) independent signals for control. In the following, we develop algorithms to construct $G_\T{max}$ and $G_\T{min}$ using the HH algorithm as an important building block. Note that here we allow self-loops in the graphical realizations. However, the designed algorithms can be adopted to the case of self-loop free networks using the machinery developed in Ref.~\cite{kim2012constructing}.

\subsection{Maximum driver node networks}\label{sec:maxprob}

Our goal is to construct $G_\T{max}$, a graphical realization of a given BDS that requires the maximum number of control signals $N^\T{max}_\T{D}$. Due to the mapping between the minimum control signal and the maximum matching problems, this is equivalent to finding a realization with the smallest maximum matching.  
We first find $N^\T{UB}$, an upper bound for $N_\T D^\T{max}$, i.e., $N^\T{UB}_\T D\geq N_\T D^\T{max}$.
Then we introduce an algorithm that aims to construct a network that achieves or approximates this bound. 
Although we do not rigorously prove that the algorithm always finds the optimum, the upper bound allows us to asses the quality of the solution:
if  $N_\T D^\T{max} = N^\T{UB}_\T D$, we know that it corresponds to the global maximum.

We start by recalling K{\H o}nig's theorem, which states that the size of the maximum matching in a bipartite network $\hat B = (V^+, V^-, \hat E)$ is equal to the size of its minimum vertex cover~\cite{konig1931graphs}. A minimum vertex cover is a minimum cardinality subset of nodes $V_\T{mvc}\subset V^+\cup V^-$, such that each link $e\in \hat E$ is adjacent to at least one node $v\in V_\T{mvc}$. Therefore to construct a network with maximum driver nodes, we need to construct a network with minimum $\lvert V_\T{mvc}\rvert$. To do this, we color a set of nodes black and the rest of the nodes white, and we then attempt to construct a graphical network such that the black nodes form a vertex cover. For the black nodes to be a vertex cover, we require that all links are adjacent to at least one black node; therefore, to minimize $\lvert V_\T{mvc}\rvert$,   our strategy is to color high-degree nodes black.

To obtain an upper bound of $N^\T{max}_\T{D}$, we sort the nodes $V^+\cup V^-$ in descending order according to their degrees and color nodes black until 
\begin{equation}
\sum_{v_i\in B^+} k^+_i + \sum_{v_i\in B^-} k^-_i \geq L \label{linkcondition}
\end{equation}
is satisfied, where $B^+$ and $B^-$ are the set of black nodes in $V^+$ and $V^-$, respectively. The remainder of the nodes are colored white. Clearly, $\lvert B^+\rvert+\lvert B^-\rvert$ is a lower bound of the minimum vertex cover for any realization of the BDS; and therefore an upper bound for the maximum number of drivers is

\begin{equation}\label{eq:ub_max}
N^\T{max}_\T{D}\leq \max(N - \lvert B^+\rvert - \lvert B^-\rvert ,1) = N^\T{UB}_\T D.
\end{equation}
If the network has a heterogeneous degree distribution, i.e., there exist hubs with a much higher number of connections than the average degree, a small number of black nodes are sufficient to satisfy the above inequality. Therefore, $N^\T{max}_\T{D}$ is expected to be high for heterogeneous degree distributions, and more restrictive for homogeneous distributions.

We now propose an algorithm that aims to construct a $\hat G_\T{max}$ such that the upper bound in Eq.~(\ref{eq:ub_max}) is achieved. The general idea is to color a set of nodes black following the same strategy as above, then search for a graphical realization such that each link is adjacent to at least one black node. If such a realization is not possible, we systematically increase the number of black nodes until we find one.

\begin{figure}[ht!]
\centering
\includegraphics[scale=0.6]{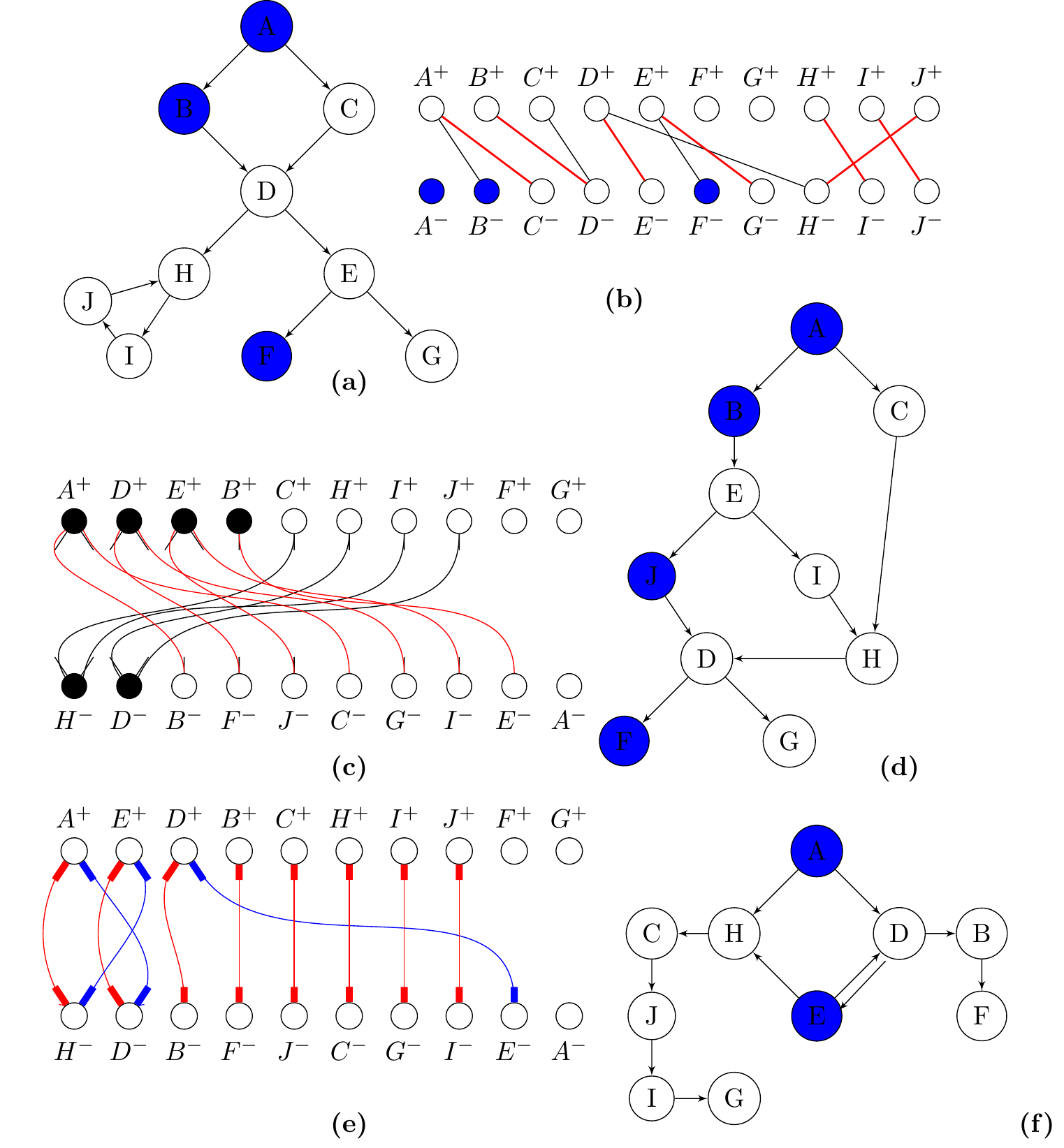}
  \caption{\label{Fig1}(a)~A network with $N=10, L=11, N_0^-=1, N_0^+=2$. (b)~The corresponding undirected bipartite representation with a maximum matching highlighted in red. Unmatched nodes (blue) on the in-side are the driver nodes. (c)~Applying Alg.~\ref{alg:max} to find $G_\T{max}$. On the out-side $A^+, D^+, E^+, B^+$ and on the in-side $D^-, H^-$ are colored black. The black nodes are reduced by non-black nodes on the other side and there remains no residual black out- and in-stubs. (d)~$G_\T{max}$ requires $N_\T{D}^\T{max}=N_\T{D}^\T{UB}=4$. (e)~Applying Alg.~\ref{alg:min} to find $G_\T{min}$. Color one stub red for the first $N-\max(N_0^+,N_0^-) = 8$ nodes on both sides and the rest of the stubs are colored blue, and we form links such that only stubs with the same color are connected. (f)~$G_\T{min}$ requires $N_\T{D}^\T{min} = N_\T{D}^\T{LB} =2$.}
\end{figure}

To check if a BDS with a given coloring $B^+$ and $B^-$ is graphical, we iteratively apply the HH algorithm. First, we reduce black nodes $B^+$ by $V^- \setminus B^-$. Note that even if the step is successful, nodes in $B^+$ may have unconnected stubs remaining. We then reduce black nodes $B^-$ by $V^+\setminus B^+$. Finally, we connect the leftover stubs by reducing the remainder of $B^+$ by the remainder of $B^-$~\footnote{Note that in the original HH algorithm without coloring, it does not matter if we reduce nodes $V^+$ by $V^-$ or $V^-$ by $V^+$. With coloring, however, we always reduce black nodes $B^+$ by non-black nodes $V^-\setminus B^+$, because the black nodes may have stubs remaining after reduction and with this order of reduction the remainder of stubs are not concentrated on a few nodes.}. If all reductions are successful, the coloring is graphical, i.e., the BDS is graphical subject to the coloring constraint, and we found $G_\T{max}$. If we are unsuccessful, we increase the number of black nodes. The pseudo-code of this algorithm is provided in Alg.~\ref{alg:max} and Fig.\ref{Fig1}(c,d) provides an example of applying the algorithm. 

The algorithm consists of repeatedly applying a modified version of the Havel-Hakimi algorithm until we find an admissible node coloring. The complexity of the Havel-Hakimi algorithm is $O(N^2\log N)$ and we test at most $N$ colorings. However, in practice the $N$ repetitions are a very conservative upper bound. Indeed, in Secs.~\ref{sec:realnets} and \ref{sec:model_nets}, we will see that for real and model complex networks, with very few exceptions, the algorithm only has to consider one candidate coloring and therefore is suitable to analyze large-scale networks. 
Furthermore, in these cases the solutions achieve the upper bound in Eq.~(\ref{eq:ub_max}), meaning that Alg.~\ref{alg:max} indeed successfully finds  $G_\T{max}$ corresponding to the global optimum. Also, note that the algorithm generates one possible $G_\T{max}$; typically, however, there are many realizations with the same number of driver nodes.  In Sec.~\ref{sec:prev_work}, we will explore such other realizations by rewiring $G_\T{max}$ such that the coloring of the nodes is respected.

\begin{algorithm}[H]

\scriptsize

\caption{\label{alg:max}Finding the graphical realization with maximum control signals for BDS $D$}
\textbf{Input:} Bi-degree Sequence $D=\{(k^+_i,k^-_i), i=1,\ldots,N\}$, \\
\textbf{Output:} Realization with maximum control signals $G_\T{max}$.
\begin{spacing}{0.9}
\begin{algorithmic}[1]
\Function{HH\_graphical}{$D,B^-, B^+$}
\If{$B^-$ is not reducible by $V^+\setminus B^+$}
\Return \textit{false}
\EndIf
\State $(B^-,V^+\setminus B^+) \gets$ reduce $B^-$ by nodes in $V^+\setminus B^+$

\If{$B^+$ is not reducible by $V^-\setminus B^-$}
\Return \textit{false}
\EndIf
\State $(B^+,V^-\setminus B^-) \gets$ reduce $B^+$ by nodes in $V^-\setminus B^-$

\If{$B^-$ is not reducible by $B^+$}
\Return \textit{false}
\EndIf

\State \Return \textit{true}

\EndFunction
\Statex
\end{algorithmic}
\begin{algorithmic}[1]
\State Sort $\mathbf{k^+}=[k^+_i], \mathbf{k^-}=[k^-_i], i=1,\ldots,N$, lists in descending order

\For{$N_B \gets 2$ to $N$ }
      \For{$N_{B^+} \gets 1$ to $N_B$}
      \State $N_{B^-} \gets N_B - N_{B^+}$
      
        \If{$\sum_{i=1}^{N_{B^+}} k^+_i + \sum_{j=1}^{N_{B^-}} k^-_j \geq L$} 
       \State $B^+=[k^+_i], i=1,\ldots,N_{B^+}, B^-=[ k^-_j], j=1,\ldots,N_{B^-} $
               
        \If{HH\_GRAPHICAL$(D,B^-, B^+)$}
        \State  $G_\T{max} \gets$ realization of $D$ with $B^-, B^+$ coloring constraint
        \State \Return $G_\T{max}$
        \EndIf
        
        \EndIf
        
\EndFor
\EndFor

\end{algorithmic}
\end{spacing}
\end{algorithm}

\subsection{Minimum driver node networks}\label{sec:minprob}

We now turn our attention to constructing $G_\T{min}$, a graphical realization of a given BDS that requires the minimum number of control signals $N^\T{min}_\T{D}$, or equivalently, the realization with the largest maximum matching. Similarly to the previous section, we first find a simple lower bound for $N^\T{min}_\T{D}$, then we introduce an algorithm that aims to construct a realization that achieves this bound.

A matching is a set of links that do not share endpoints; therefore, in a bipartite network $\hat G=(V^+,V^-,\hat E)$ a matching cannot be larger than the number of nodes with non-zero degree in $V^+$ or in $V^-$. This means that a lower bound for the minimum number of drivers is

\begin{equation}
\label{eq:lb_min}
N^\T{min}_\T{D}\geq  \max\big(N^+_0, N^-_0, 1\big) = N^\T{LB}_\T D,
\end{equation}
where $N^+_0$ is the number of sinks, i.e., nodes with zero out-degree, and $N^-_0$ is the number of sources, i.e., nodes with zero in-degree. 
Therefore, networks with a high number of sources and sinks are expected to restrict the possible number of driver nodes more, i.e., have high $N^\T{min}_\T{D}$.
 
We now propose an algorithm that aims to construct a $\hat G_\T{min}$ such that this lower bound is achieved. We again start with two sets of $N$ unconnected nodes, $V^+$ and $V^-$, and each node is assigned stubs corresponding to their prescribed degrees. Before attempting to connect the stubs, we arrange the nodes on both sides in descending order according to their degrees, and we color one of the stubs red of the first $N-\max(N^+_0, N^-_0)$ nodes on both sides, and the remainder of the stubs are colored blue. Now we sequentially form links such that only stubs of the same color are allowed to be connected to each other. If a graphical realization exists that satisfies this additional constraint, the set of red links form a matching corresponding to the bound in Eq.~(\ref{eq:lb_min}). If such a realization is not possible, we systematically lower the number of red stubs, until we find one.

To check if a BDS with coloring is graphical, we modify the HH algorithm. In each step, we pick a node $v^+_i\in V^+$. If node $v^+_i$ has a red stub, connect the red stub to the node in $V^-$ that has the most unconnected stubs and has an available red stub. Then we connect its blue stubs to the nodes in $V^-$ with the most unconnected stubs. We repeat this step for all nodes in $V^+$. If we successfully connect all stubs, the BDS with coloring is graphical. If at any step, we run out of available nodes or if at the end we have unconnected stubs left over, then we failed to find a graphical realization with the current coloring and we reduce the number of red stubs.
For this, we pick the red stub that belongs to the node with the smallest possible degree in both $V^+$ and $V^-$, and we change its color to blue. We repeat this until we find a graphical coloring. Following this strategy, we have to check at most $N - \max(N^+_0, N^-_0)\leq N$ candidate colorings. We provide the pseudo-code for the process in Alg.~\ref{alg:min} and a small example in Fig.~\ref{Fig1}e,f. 

In Secs.~\ref{sec:realnets} and \ref{sec:model_nets}, we apply Alg.~\ref{alg:min} to the BDS of a collection of real complex networks and power-law distributed model networks. We find that our algorithm achieves the lower bound given by Eq.~(\ref{eq:lb_min}) for all real networks and all model networks with sensibly chosen parameters. This means that (i)~in these cases our algorithm indeed successfully finds the $G_\T{min}$ corresponding to the global optimum, (ii)~the algorithm only has to consider one candidate coloring and therefore it terminates quickly, making it suitable to analyze large-scale complex networks.

Note that this algorithm, similar to the Alg.~\ref{alg:max}, generates one possible $G_\T{min}$; typically, however, there are many realizations with the same number of driver nodes. In Sec.~\ref{sec:prev_work}, we will explore other realizations by rewiring $G_\T{min}$ such that the coloring of the nodes is respected.

\begin{algorithm}[H]
\scriptsize
\caption{\label{alg:min}Finding the graphical realization with minimum required control signals for BDS $D$}
\textbf{Input:} Bi-degree Sequence $D=\{( k^+_i,  k^-_i), i=1,\ldots,N\}$, \\
\textbf{Output:} Realization with minimum control signals $G_\T{min}$.

\begin{spacing}{1}
\begin{algorithmic}[1]
\Function{HH\_graphical}{$D, M$}

\State Sort $V^+, V^-$ lists in descending order based on their degrees

\State  Color one stub of the first $M$ elements of $V^+, V^-$ as red and the remaining by blue 

\For{each node $v^+$ }
\If{$v^+$ has red out-stub}
\State reduce the first red-stub in $V^-$ about $v^+$  
\EndIf
\If{$v^+$ has blue stubs}
\If{blue stubs of $V^-$ is reducible about blue stubs of $v^+$}
\State reduce the blue stubs of $V^-$ about the blue stubs of $v+$
\Else
\State \Return \textit{false}
\EndIf
\EndIf
\EndFor
\State \Return \textit{true}

\EndFunction
\Statex
\end{algorithmic}

\begin{algorithmic}[1]
\State $N^+_0 \gets$ number of sinks, $N^-_0 \gets $ number of sources
\State $M \gets N - \max\{N^+_0,N^-_0\}$
    
\While{not\big(HH\_GRAPHICAL($D,M$)\big)}
\State $M \gets M - 1$

\State Color one of the stubs red of the first $M$ nodes on both sides 
\State $G_\T{min} \gets$ realization of $D$ with coloring constraint   
 
\EndWhile

\State \Return $G_\T{min}$

\end{algorithmic}
\end{spacing}
\end{algorithm}

\section{Real networks}\label{sec:realnets}

\begin{figure*}[ht!]
\centering
\includegraphics[ scale=0.6]{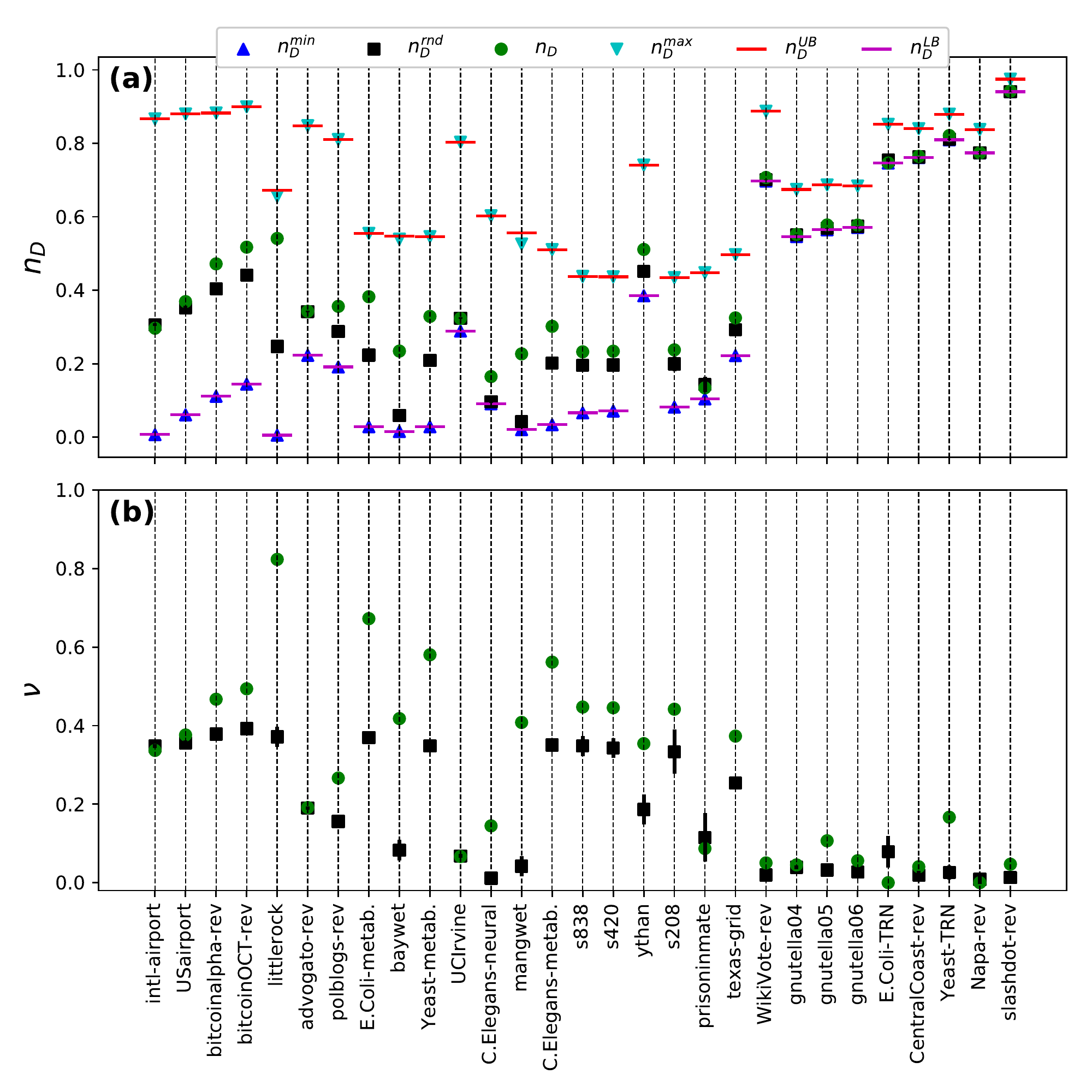}
  \caption{(a)~$ n_\T D, n_\T D^{\T{min}},n_\T D^{\T{max}}$, and $n_\T D^{\T{rnd}}$ for different real networks and their corresponding bounds. (b)~The \bl{structural control} complexity $\nu$ for original and randomized networks. In (a) and (b) the results for the random case are the average of 20 randomizations and the error bars represent the standard deviation, which may be smaller than the marker sizes.}\label{Fig2}
\end{figure*}

We now apply our algorithms to analyze a collection of real networks. For each network, we calculate $N_\T D$, the number of driver nodes necessary to control the original network. We then measure $N_\T{D}^\T{max}$ and $N_\T{D}^\T{min}$ together with their corresponding upper and lower bounds provided by Eqs.~(\ref{eq:ub_max}) and (\ref{eq:lb_min}). Finally, we randomize each network preserving their degree distribution and measure $N_\T{D}^\T{rnd}$, the number of drivers averaged over 20 independent randomizations. We summarize these results along with descriptions of the datasets in Table~1 of the supplementary material.

Figure~\ref{Fig2}(a) shows the results of the measurements for each network normalized by the total number of nodes. The first notable observation is that for all real BDSs Alg.~\ref{alg:min} finds graphical realizations such that $n_\T D^\T{min}=N_\T D^\T{min}/N$ reaches its lower bound; and Alg.~\ref{alg:max} finds realizations such that $n_\T D^\T{max}=N_\T D^\T{max}/N$ is equal to its predicted upper bound for all but three exceptions. These exceptions are three of the four food webs: \emph{mangwet}, \emph{baywet}, and  \emph{littlerock}. As we will discuss more in the next section, the reason why $n_\T D^\T{max}$ does not reach its bound in these cases is rooted in their small size and their very broad in- or out-degree distributions. 

The number of driver nodes needed to control a graphical realization of a BDS is restricted to the range $[n_D^{\T{min}}, n_D^{\T{max}}]$.
In Fig.~\ref{Fig2}, we ordered the networks according to $n_D^{\T{max}} - n_D^{\T{min}}$ such that networks with the least restrictive BDS are to the left, and the most restrictive ones are to the right. For some networks, like the airline or bitcoin networks, there both exist realizations such that we need considerably more or considerably fewer  driver nodes than the original networks or their randomized counterparts. While other networks, like gene transcription networks and some trust networks, are characterized by a very narrow range $[n_D^{\T{min}}, n_D^{\T{max}}]$; and therefore their BDS largely determines $n_\T D$. 

To understand which features of the BDS affect the range of possible $n_\T D$ values, recall that in Sec.~\ref{sec:maxminprob} we showed that $N^\T{min}_\T D$ increases with increasing number of sources or sinks, and $N^\T{max}_\T D$ is high for networks with large hubs. Sparse networks with low average degree typically have many sources and sinks, which narrows the range $[n_D^{\T{min}}, n_D^{\T{max}}]$. The role of heterogeneous degree distribution is less straight forward: heterogeneous networks have large hubs, increasing $n_D^{\T{max}}$; however, heterogeneous networks also typically have more sources and sinks than homogeneous networks with the same average degree, increasing $n_D^{\T{min}}$. Ultimately, the balance of these two features determines the net effect of degree heterogeneity on $[n_D^{\T{min}}, n_D^{\T{max}}]$.


The fraction of driver nodes needed to control a network is determined by the degree sequence of the nodes and how these nodes are connected to each other. To quantify the effect of network structure beyond the degree sequence, we introduce the \bl{structural control} complexity of a network as 
\begin{equation}\label{eq:contrcomp}
\nu = \frac{N_D-N^\T{min}_\T{D}}{N^\T{max}_\T{D}-N^\T{min}_\T{D}}.
\end{equation}
\bl{Structural control} complexity is normalized such that $\nu=1$ if the network is the hardest and $\nu=0$ if it is the easiest to control given its BDS; high $\nu$ values indicate richer internal structure with respect to controllability. Previously, we showed that $N_\T D^\T{max}$ and $N_\T D^\T{min}$ are equal to their upper and lower bounds for the exceeding majority of networks;  we can, therefore, accurately estimate $\nu$ using these bounds. However, to generate a graphical realization with $N_\T D^\T{max}$ or $N_\T D^\T{min}$ driver nodes, we still have to use Algs.~\ref{alg:max} and \ref{alg:min}.

On Fig.~\ref{Fig2}(b), we show $\nu$ for the collection of real networks. Strikingly, networks that are considered hard to control with respect to $n_\T D$, such as transcription, p2p, and some trust networks, have very low \bl{structural control} complexity. This means that most driver nodes in these networks are either sources or sinks.
Other networks, such as electronic circuits and food webs, have higher $\nu$ values than $n_\T D$, meaning that the complexity of their structure beyond their BDS is what makes these networks hard to control.

Interestingly, we find two types of trust networks: some have high $n_\T D$ and low $\nu$ (\emph{Slashdot}, \emph{Napa}, \emph{Central Coast}, and \emph{WikiVote}) and others have comparable $n_\T D$ and $\nu$ (\emph{Prison Inmate}, \emph{Advogato}, \emph{BitcoinOCT}, and \emph{BitcoinAlpha}). In trust networks a link from individual $a$ to $b$ indicates that $b$ trusts or seeks advice from $a$. 
Networks in the first group rely on a few highly trusted actors, i.e., the network structure is dominated by star-like patterns. For example, Ref.~\cite{levy2018innovation} showed that the networks representing influence between viticulture growers (\emph{Central Coast} and \emph{Napa}) are centered around a few disproportionately influential actors.
On the other hand,  trust networks in the second group have less centralized structure. For example, in the networks extracted from online platforms where Bitcoin users vouch for their peers (\emph{BitcoineAlpha} and \emph{BitcoinOCT}) no central authority exists, trust is distributed and encoded in the network structure.

\begin{figure}[ht!]
\centering
\includegraphics[scale=0.45]{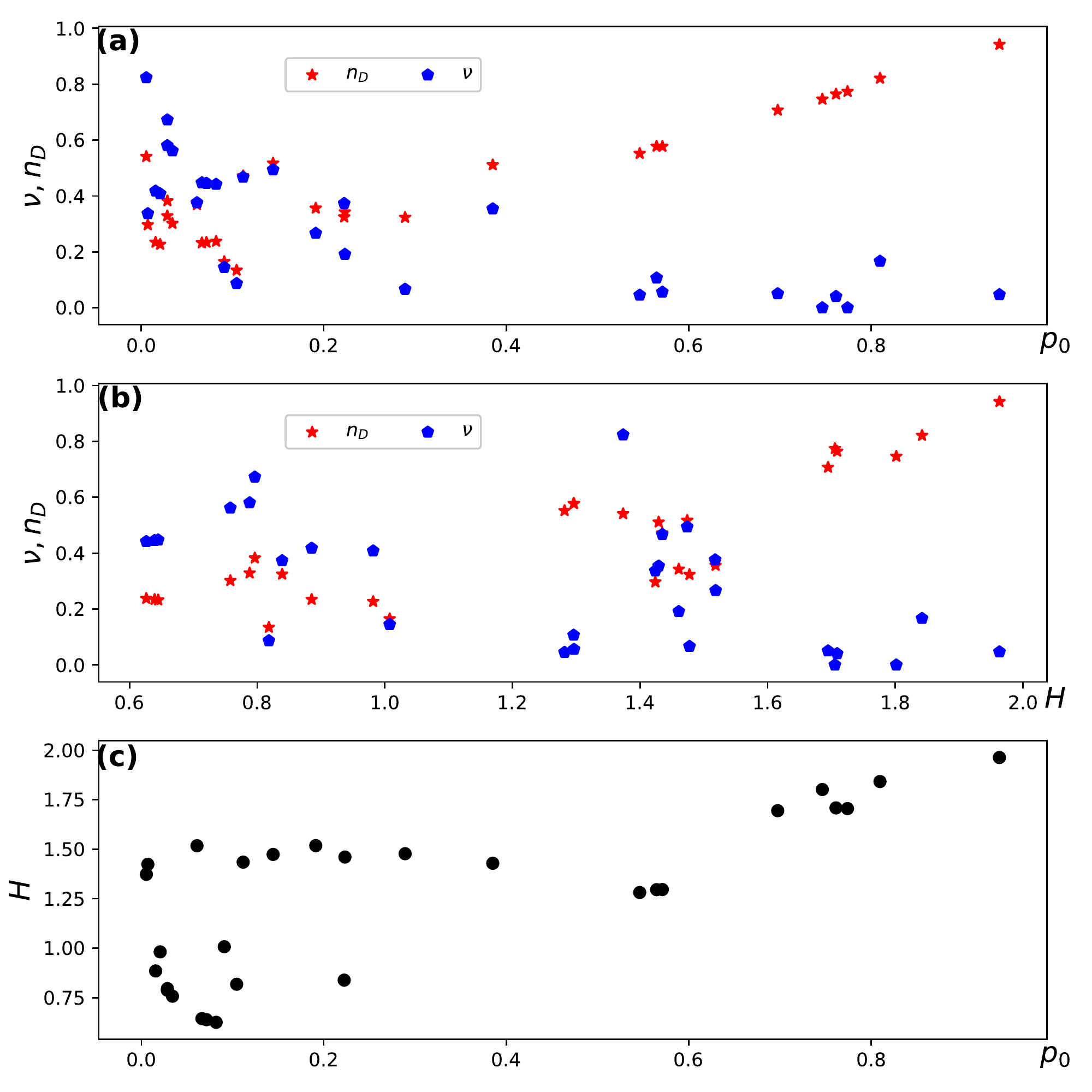}
  \caption{\label{fig:heterogeneity} (a)~Scatter plot showing the relationship between $p_0=\max\{N^+_0,N^-_0\}/N$ and $n_\T D$ and $\nu$ for real networks. Increasing the number of sources and sinks increases the number of driver nodes, yet typically decreases \bl{structural control} complexity. (b)~The relationship between $H=\max\{H^+, H^-\}$ and $n_\T D$ and $\nu$ shows similar pattern. (c)~The similar relationship is explained by the positive correlation between $H$ and $p_0$.}
\end{figure}

Finally, on Figs.~\ref{fig:heterogeneity}a,b we plot $\nu$ against the maximum of the fraction of sources or sinks $p_0=\max\{N^+_0,N^-_0\}/N$ and the degree heterogeneity $H=\max\{H^+, H^-\}$, where
\begin{equation}
H^{+/-} = \frac{1}{c N^2}\sum\limits_{i,j=	1}^N \lvert k^{+/-}_i - k^{+/-}_j \rvert
\end{equation} 
and $k^{+/-}_i$ is the out- or the in-degree of node $i$~\cite{liu2011controllability}.
We find negative correlation between $\nu$ and $p_0$, i.e., the presence of sources and sinks typically reduces \bl{structural control} complexity. Networks with a surprising amount of sinks, such as \emph{Slashdot}, have very low corresponding \bl{structural control} complexity. Figure~\ref{fig:heterogeneity}b shows a similar negative correlation between $\nu$ and $H$; this relationship, however, has to be interpreted carefully.
In the next section, we will see that increasing degree heterogeneity in model networks in fact may increase  $\nu$.
The apparent contradiction is resolved noticing $p_0$ and $H$ are not independent quantities: heterogeneous networks have more sources and sinks than homogeneous networks with the same average degree. Indeed, we observe a positive correlation between $p_0$ and $H$ for the real networks (Fig.~\ref{fig:heterogeneity}c).

\section{Model networks}\label{sec:model_nets}

In this section, we systematically investigate how various characteristics of degree distributions affect $n_\T D^\T{max}$, $n_\T D^\T{min}$, and the \bl{structural control} complexity $\nu$ using model networks. Also, we demonstrate through simulations and analytical arguments that $n_\T D^\T{max}$ and $n_\T D^\T{min}$ are equal to their corresponding upper and lower bounds for typical degree sequences, except for very small and very heterogeneous networks.

For our investigations, we need to generate degree sequences (i)~that are drawn from a tunable degree distribution, we are particularly interested in degree heterogeneity and the presence of sources and sinks, and (ii)~that are graphical even for small samples and highly heterogeneous degrees.
To achieve this, we use a generalized version of the directed static model~\cite{liu2011controllability,goh2001universal,chung2002connected}.
To simplify notation, we only consider networks with symmetric out- and in-degree distribution, i.e., $p_\T{out}(k)\equiv p_\T{in}(k)\equiv p(k)$; however, all of our results are easily extended to the general case.
To generate a network, we start with $N$ unconnected nodes, and we assign a weight $w_i=i^{-\alpha}/(\sum_j j^{-\alpha})$ to nodes $i=1,2,\ldots,N-N_0$ where $\alpha \in [0,1)$, and $w_i=0$ to nodes $i=(N-N_0+1),\ldots,N$.
 We then randomly select two nodes $i$ and $j$ with probability $w_i$ and $w_j$, respectively, and if there is no directed link from node $i$ to $j$, we connect them. We repeat this step until $L$ links are added. Setting $N_0=0$ we obtain the original static model~\cite{goh2001universal,chung2002connected}. 
 
The resulting network has average in- and out-degree $c=L/N$, and both its in- and out-degree distributions can be approximated as a sum of binomial distributions
\begin{equation}\label{eq:pk}
p(k) = \frac{1}{N} \sum_{i=1}^{N} \binom{L}{k} w_i^k (1 - w_i)^{L-k}.
\end{equation}
For large $N$, approximating the binomial distributions with Poisson distributions and substituting integrals for summations~\cite{catanzaro2005analytic}, we obtain
\begin{equation}\label{eq:pkapprox}
\begin{split}
p(k)=n_0\delta_{0,k}+& (1-n_0)\frac{(c[1-\alpha]/[1-n_0])^{1/\alpha}}{\alpha}\times \\
\times &\frac{\Gamma(k-1/\alpha,c[1-\alpha]/[1-n_0])}{\Gamma(k+1)},
\end{split}
\end{equation}
where $n_0=N_0/N$, $\delta_{j,k}$ is the Kronecker delta, $\Gamma(z)$ is the gamma function, and $\Gamma(z,a)$ is the upper incomplete gamma function. The tail of the distribution decays  as a power-law, i.e., 
\begin{equation}\label{eq:pkasymp}
p(k) \simeq (1-n_0)\frac{\left(c[1-\alpha]/[1-n_0]\right)^{1/\alpha}}{\alpha} k^{-(1+1/\alpha)} \sim k^{-\gamma},
\end{equation}
where $\gamma = 1+ 1/\alpha$ is the degree exponent.


The expected degree of node $i$ is $c_i= w_i L$; and $i=1$ provides the expected maximum degree in the network: 
\begin{equation}\label{eq:kmax}
k_\T{max} = \frac{L}{\sum_{j=1}^{N-N_0}j^{-\alpha}} \approx \frac{c}{1-n_0}(1 - \alpha)(N-N_0)^\alpha,
\end{equation}

Note that alternatively, we could use the configuration model to generate networks, which would allow us to directly choose the degree distribution $p(k)$. The advantage of the static model is that it always generates a graphical degree sequence, while in the case of the configuration model, this becomes increasingly difficult for heterogeneous degree distributions, i.e., as degree exponent $\gamma$ approaches 2. The disadvantage of the static model, however, is that for $\gamma<3$, the expected number of links between some node pairs becomes greater than one. Since multiple links are not allowed, excess links are rewired; and Eq.~(\ref{eq:pk}) becomes only approximate. This effect is the strongest for degree exponents close to 2, but the correction becomes less pronounced as network size increases. \bl{See supplementary material for more details about the derivations and approximations of (\ref{eq:pk}) to (\ref{eq:pkasymp}).}

\subsection{Maximum and minimum driver nodes}

We now investigate how average degree, degree heterogeneity, and the fraction of sources and sinks affect $n_\T D^\T{max}$, $n_\T D^\T{min}$, and $\nu$. In addition to numerical measurements, we also provide analytical formulas for the upper bound of $n_\T D^\T{max}$ and the lower bound of $n_\T D^\T{min}$.

In Sec.~\ref{sec:minprob}, we showed that the lower bound of the number of driver nodes is simply the maximum of the number of sources or sinks. Therefore, following Eqs.~(\ref{eq:pk}) and (\ref{eq:pkapprox}) we get
\begin{align}
\begin{split}\label{eq:LB_anal}
n^\T{LB}_\T D &= p(0) = n_0+(1-n_0)\sum_{i=1}^{N-N_0}e^{-cNw_i}\approx\\
&\approx n_0 + (1-n_0)\frac{(c[1-\alpha]/[1-n_0])^{1/\alpha}}{\alpha}\times \\
&\times \Gamma(-1/\alpha,c[1-\alpha]/[1-n_0]).
\end{split}
\end{align}
Expanding for large average degree, we get $n^\T{LB}_\T D-n_0\sim\Exp(-Ac)c^{-\gamma}$, where $A=(1-\alpha)/(1-n_0)$.

Obtaining the upper bound $n^\T{UB}_\T D$ is less straight forward. Recall that in Sec.~\ref{sec:maxprob}, we derived the upper bound by coloring the highest degree nodes black in the bipartite representation of the network until the number of links adjacent to black nodes was at least $L$, and the upper bound is then $N^\T{UB}_\T D=N-\lvert B^+\rvert-\lvert B^+\rvert$ where $B^+$ and $B^-$ are the sets of black nodes in the two sides of the bipartite representation. Since we only consider symmetric degree distributions, we can write
\begin{equation}\label{eq:UB_anal}
n^\T{UB}_\T D=1-2n_\T B,
\end{equation}
where $n_\T B$ is the expected fraction of black nodes on either side of the bipartite network. We color the nodes black starting from the highest degree, meaning that we color all nodes black with degree larger than some degree $k_0$ and a $q$ fraction of nodes that have degree $k_0$. We require that at least half of the links must be adjacent to black nodes; therefore, for large $N$, $k_0$ and $q$ must satisfy the equation
\begin{equation}\label{eq:k0}
k_0 q + \sum_{k=k_0+1}^{\infty} kp(k)  = 0.5\,c.
\end{equation}

 \bl{Since the binomial approximation of the degree distribution in Eq. (\ref{eq:pk}) is more accurate for low $k$, we numerically solve $k_0 q^\prime  + \sum_{k=1}^{k_0-1} kp(k)  = 0.5\,c$; noting that the other half of the links are connected to non-black nodes.} Having $k_0$ and $q^\prime$, we get $n_\T B$ 

\begin{equation}\label{eq:nB}
n_\T B= q+\sum_{k=k_0 + 1}^\infty p(k) = 1 - \left(q^\prime + \sum_{k=0}^{k_0 -1}p(k)\right).
\end{equation}
We can also obtain an approximate closed-form solution by using the asymptotic form of $p(k)$ provided in Eq.~(\ref{eq:pkasymp}) and substituting the summation with an integral:
\begin{equation}
(1-n_0)\int_{k_0}^\infty \frac{\left[c(1-\alpha)/(1-n_0)\right]^{\gamma-1}}{\alpha} k^{-\gamma+1}dk=0.5c.
\end{equation} 
Solving the above equation provides $k_0$, which we can use to calculate the fraction of black nodes 

\begin{equation}\label{eq:black_approx}
n_\T B \approx \int_{k_0}^\infty p(k)dk = \left[1-n_0\right]2^{-\frac{\gamma-1}{\gamma-2}}.
\end{equation}

This indicates that as $\gamma$ approaches 2, a vanishing fraction of nodes will be colored black; and therefore $n^\T{UB}_\T D$ approaches 1. Notably, the solution does not depend on the average degree; we have to consider, however, that this solution is expected to well approximate the exact solution for homogeneous and dense networks. \bl{See supplementary material for more details about the derivations and approximations of equations in this subsection.}

\begin{figure}[ht!]
\centering
\includegraphics[scale=0.5]{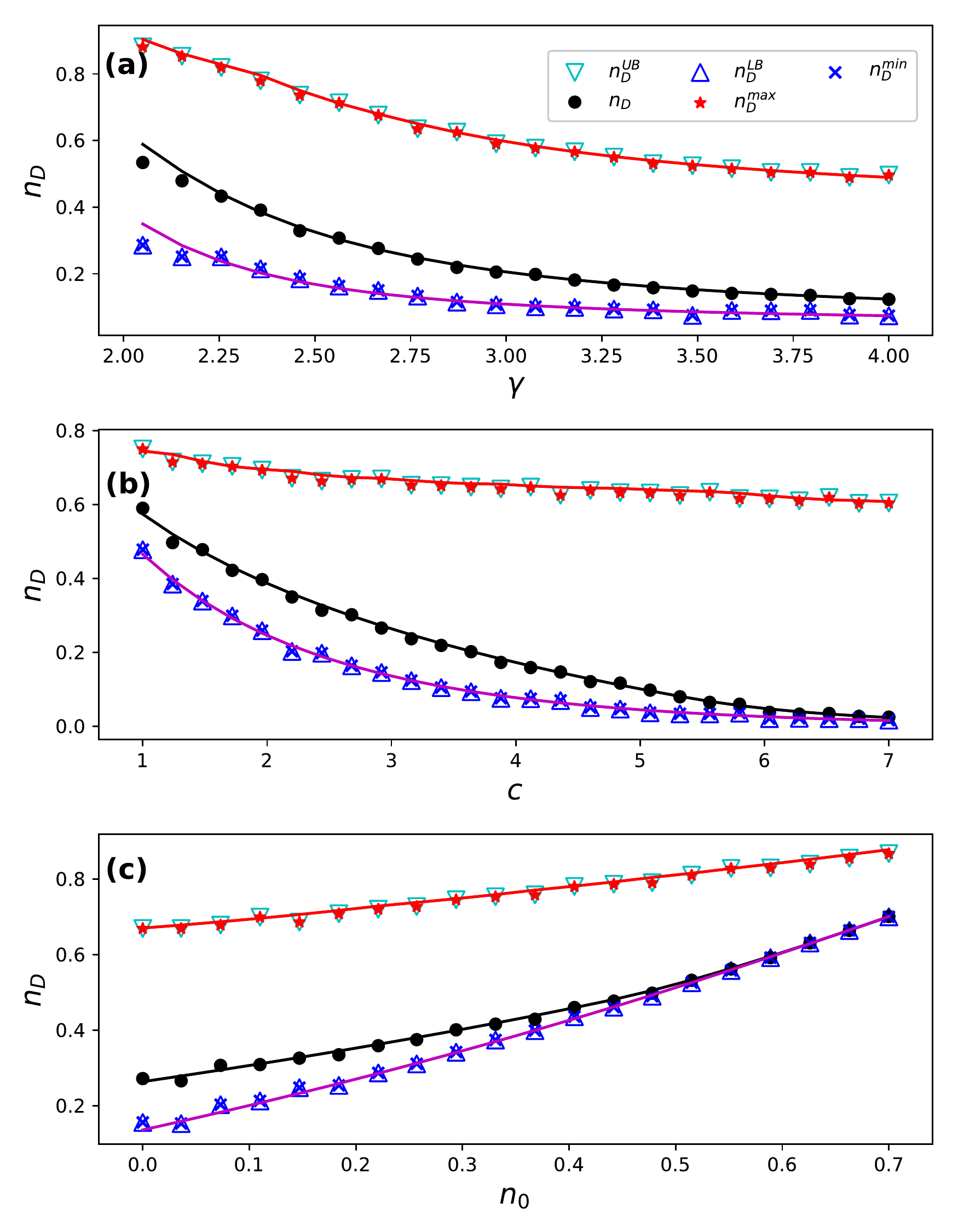}
  \caption{\label{fig:model_minmax}(a) $n_D$ for power-law networks generated by static model with $N=1000, N_0=0$ nodes and $E=3000$ links for different values of $\gamma$. (b) $n_D$ for model networks with $N=1000, N_0=0, \gamma=2.7$ where the average degree is changing from $1$ to $7$. (c) $n_D$ for model networks with $N=1000, \gamma=2.7$ where $N_0$ is changing from $0$ to $0.7N$. Numerical results are the average of 20 independent networks, and the error bars indicate the standard deviation, typically smaller than the marker size.}
\end{figure}

Figure~\ref{fig:model_minmax} shows results for model networks. We generate instances of BDSs and we calculate $n_\T D$, $n^\T{min}_\T D$, $n^\T{max}_\T D$, $n^\T{LB}_\T D$, and $n^\T{UB}_\T D$, while systematically changing the degree exponent $\gamma$, the average degree $c$, and the parameter $n_0$.
The symbols are numerical measurements and continuous lines are analytical results. The numerical measurements and theory are in great agreement, except for very heterogeneous degree distributions where finite size effects become non-negligible.
We observe that similarly to real networks, the lower bound $n^\T{LB}_\T D$ and the upper bound $n^\T{UB}_\T D$ are exactly equal to the $n^\T{min}_\T D$ and $n^\T{max}_\T D$, respectively. In the next section, we will show that the lower and upper bounds are not reached only for very small and heterogeneous networks.

The minimum fraction of driver nodes $n^\T{min}_\T D$ is equal to the fraction of zero-degree nodes, the parameter $n_0$ tunes the number of sources and sinks; therefore, increasing $n_0$ increases $n^\T{min}_\T D$ (Fig.~\ref{fig:model_minmax}c). Decreasing the average degree $c$ or the degree exponent $\gamma$ indirectly increase the number of sources and sinks, therefore also increase $n^\T{min}_\T D$ (Figs.~\ref{fig:model_minmax}a-b). 
The maximum fraction of driver nodes  $n^\T{max}_\T D$ depends on high-degree nodes; therefore the degree exponent $\gamma$ has the strongest, while $c$ and $n_0$ have a weaker effect on $n^\T{max}_\T D$. In fact, we showed in Eq.~(\ref{eq:black_approx}) that $n^\T{UB}_\T D$, and  therefore $n^\T{max}_\T D$, is approximately independent of $c$, which is supported by numerical results (Fig.~\ref{fig:model_minmax}b). 
The fraction of driver nodes $n_\T D$ for the static model is typically closer to its  minimum than its maximum; and approaches $n^\T{min}_\T D$ as $\gamma$, $c$, or $n_0$ increases.

\begin{figure*}[ht!]
\centering
\includegraphics[scale=0.53]{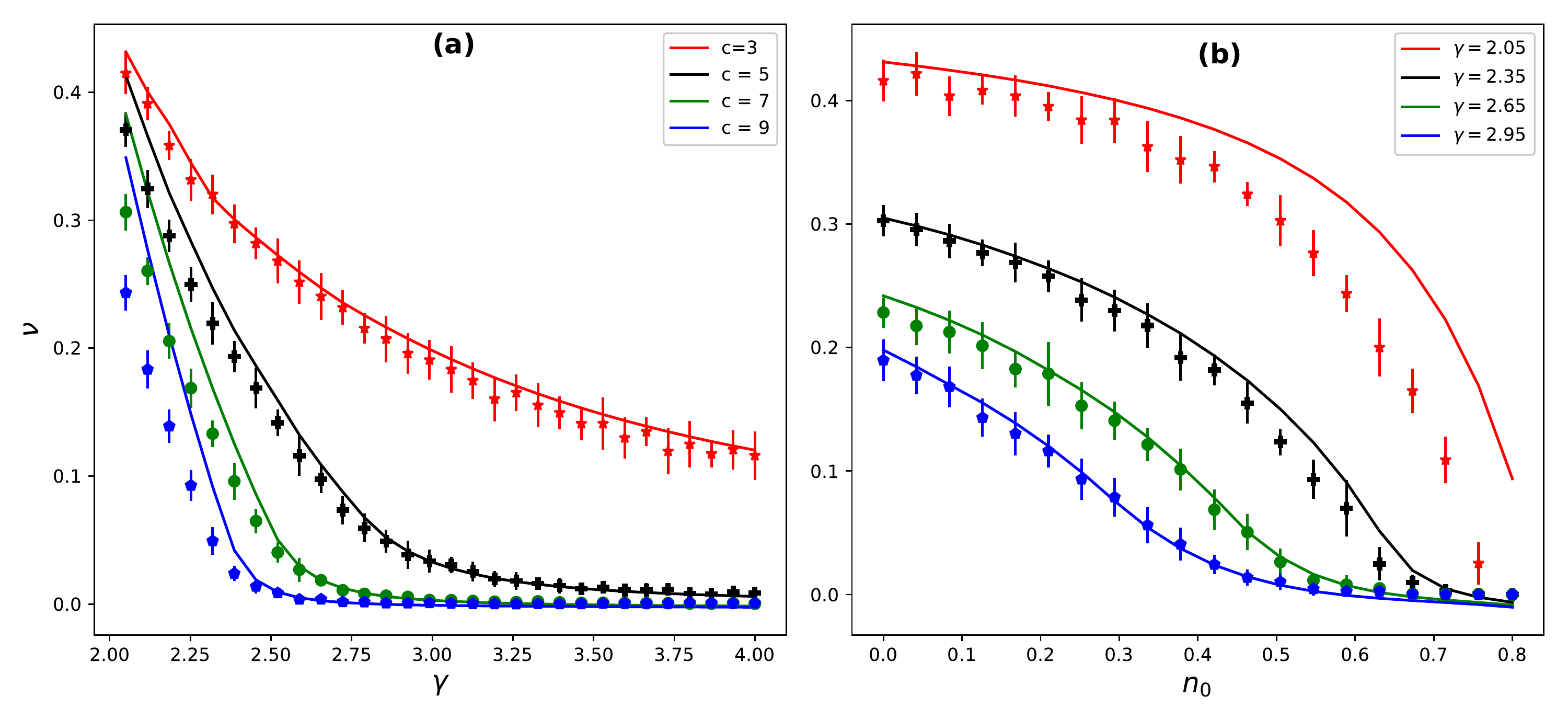} 
  \caption{\label{fig:model_nu}(a)~\bl{Structural control} complexity $\nu$ as a function of degree exponent $\gamma$ for different values of average degree $c$, (b)~$\nu$ as a function of $n_0$ for different values of $\gamma$. Lines provide the analytical solution, symbols show numerical results for an average of 20 independent networks with $N=1000$ nodes, and error bars indicate the standard deviation.}
\end{figure*}

Next, we examine the \bl{structural control} complexity $\nu$ of model networks. Figure~\ref{fig:model_nu} shows $\nu$ as a function of the degree exponent $\gamma$ and the parameter $n_0$ for various values of average degree $c$.
The continuous lines represent analytical results; since we found that $n^\T{min}_\T D=n^\T{LB}_\T D$ and $n^\T{max}_\T D=n^\T{UB}_\T D$, we calculate $\nu$ using the analytical solution of $n^\T{LB}_\T D$ and $n^\T{UB}_\T D$ provided in Eqs.~(\ref{eq:LB_anal}) and (\ref{eq:UB_anal}), respectively.
We find that similarly to the fraction of driver nodes $n_\T D$~\cite{liu2011controllability}, \bl{structural control} complexity $\nu$ is increased by degree heterogeneity; however, unlike $n_\T D$, increasing average degree $c$ decreases $\nu$.
We also increase the fraction of sources and sinks by increasing the parameter $n_0$, Fig.~\ref{fig:model_nu}b shows that this leads to low $\nu$ even for very heterogeneous networks. This is in line with what we found for real networks: in Fig.~\ref{Fig2}, we found a group of networks that are characterized by high $n_\T D$ yet low $\nu$, and these networks had a very high fraction of sources or sinks.

\subsection{Effectiveness of the upper and lower bound}

\bl{We previously shown that $n^\T{UB}_\T D$  and $n^\T{LB}_\T D$ rigoroulsy provide bounds for the number of driver nodes. Applying Alg.~1 and 2 to degree sequences of real and realistic model networks, we observed that the maximum fraction of driver nodes $n^\T{max}_\T D$ is equal to its upper bound $n^\T{UB}_\T D$ (with the exception of a few food webs) and that the minimum fraction of driver nodes $n^\T{min}_\T D$ is equal to its lower bound $n^\T{LB}_\T D$ always. This prompts the question, what properties of the degree sequences allow $n^\T{UB}_\T D$  and $n^\T{LB}_\T D$ to be tight bounds?} In this section, we use numerical simulations and analytical arguments to show that $n^\T{max}_\T D$ and $n^\T{min}_\T D$ are not equal to their respective bounds only for very small and heterogeneous networks.

\begin{figure*}[ht!]
\centering
\includegraphics[scale=0.51]{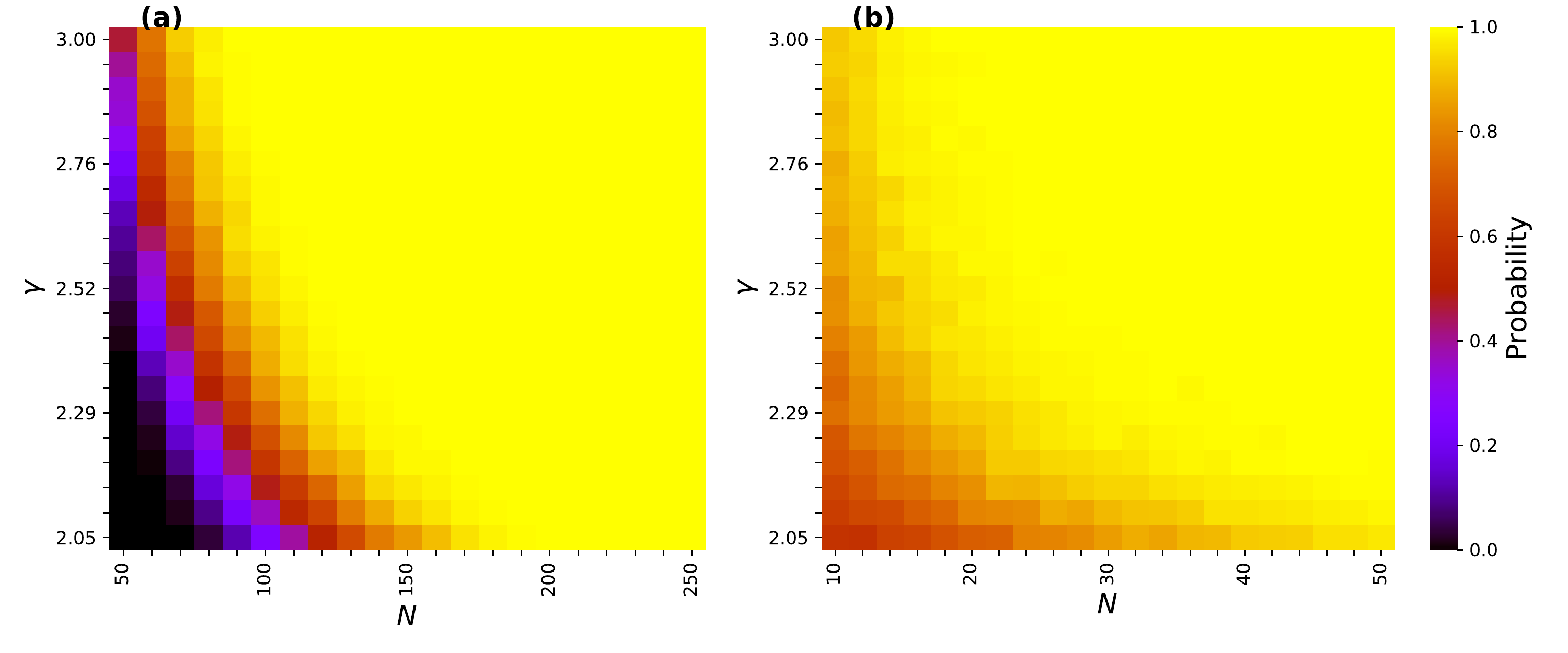}
  \caption{\label{fig:low-upp-heatmap}(a)~The probability of $n^\T{max}_\T D= n^\T{UB}_\T D$ for networks with $c=15$ generated using the static model. (b)~The probability of $n^\T{min}_\T D= n^\T{LB}_\T D$ for networks with $c=2$ generated using the static model. The probabilities were estimated using 1000 independent realizations.}
\end{figure*}

We generate BDSs while systematically changing the number of nodes $N$ and the degree exponent $\gamma$, and we compare the $n^\T{max}_\T D$ to its corresponding upper bound $n^\T{UB}_\T D$. Figure~\ref{fig:low-upp-heatmap}a shows the probability that $n^\T{max}_\T D= n^\T{UB}_\T D$, we find that the chance that equality does not hold increases for very small or very heterogeneous networks. Finding $n^\T{max}_\T D\neq n^\T{UB}_\T D$ means that Alg.~\ref{alg:max} fails when attempting to create a graphical realization corresponding to the equality. Recall that the algorithm works by taking the bipartite representation of the BDS and coloring a set of the highest degree nodes black such that $N^\T{UB}_\T D=N-\lvert B^+\rvert-\lvert B^+\rvert$ where $B^+$ and $B^-$ are the sets of black nodes on the two sides of the bipartite representation, the rest of the nodes are colored white. Then it attempts to create a graphical realization with the additional requirement that each link is adjacent to at least one black node.
 Through numerical investigations of the static model, we found that if $n^\T{max}_\T D\neq n^\T{UB}_\T D$, Alg.~\ref{alg:max} fails at connecting the highest degree non-black node. Using Eq.~(\ref{eq:k0}), we found for large enough networks that the maximum degree of non-black nodes $k_0$ is constant, e.g., it becomes independent of $N$. To connect this non-black node without creating double links, we require at least $k_0$ black nodes on the other side of the bipartite BDS. On the other hand, using Eq.~(\ref{eq:nB}), we found that the number of black nodes is
\begin{equation}
N_\T B\propto N.
\end{equation}
The algorithm fails if $k_0>N_\T B$; since $k_0$ does not depend on system size, we expect this to happen only for small networks, and $n^\T{max}_\T D = n^\T{UB}_\T D$ holds true for large networks always.


Figure~\ref{fig:low-upp-heatmap}b shows the probability that $n^\T{min}_\T D= n^\T{LB}_\T D$. We find that it is even less likely that equality does not hold than for the maximum driver node case and the region where it happens is restricted to even smaller and more heterogeneous networks. The fact that $n^\T{min}_\T D~\neq~n^\T{LB}_\T D$ means that Alg.~\ref{alg:min} fails when attempting to wire a graphical realization corresponding to the lower bound. Recall that Alg.~\ref{alg:min} does this by taking the bipartite representation of the BDS, and coloring one stub red of each of the $N-\max(N^+_0,N^-_0)$ highest degree nodes while coloring the rest of the stubs blue, where $N^+_0$, $N^-_0$ is the number of nodes with zero-degree on the two sides of the BDS. The algorithm then attempts to connect the blue stubs to blue stubs and red stubs to red stubs. Through our numerical measurements, we found that when $n^\T{min}_\T D\neq n^\T{LB}_\T D$, the algorithm most frequently fails at connecting the blue stubs of the highest degree node. Following Eq.~(\ref{eq:kmax}), the highest expected degree is
\begin{equation}
k^\T B_\T{max} = k_\T{max}-1\propto N^\alpha,
\end{equation}
for large networks. To connect these stubs without creating double links, we require at least as many nodes with nonzero blue stubs on the other side of the bipartite representation. In model networks with symmetric in- and out-degree distributions, we color one stub red for each node; therefore the number of nodes with at least one blue stub is
\begin{equation}
N^\T B_\T{nz} = (1-p(0)-p(1))N.
\end{equation}
The algorithm fails if $k^\T B_\T{max}>N^\T B_\T{nz}$; since $\alpha < 1$ we only expect this to happen for very small networks, and $n^\T{min}_\T D = n^\T{LB}_\T D$ holds true for large networks always.

\section{Relation to previous work}\label{sec:prev_work}

In this section, we discuss the relationship between our work and previous work on network control, and we show that understanding how the degree sequence constrains the number of driver nodes allows us to better interpret established results. We focus on two main findings, the role of degree correlations between connected node pairs and the so-called control profile of complex networks~\cite{posfai2013effect,ruths2014control}.

\subsection{Degree correlations}

Reference~\cite{posfai2013effect} investigated how higher order structural features beyond the degree distribution affect the fraction of driver nodes $n_\T D$ needed to control complex networks. The authors found that out-in degree correlations, i.e., correlations between the out-degree of the source node and the in-degree of the target node at the two ends of a directed link, have a strong effect on $n_\T D$.
Specifically, they added correlations via randomized rewiring of networks while keeping the degree distribution fixed, and they showed that negative out-in degree correlation increased, while positive correlation decreased $n_\T D$. Here, we explore a complementary question: what structural patterns characterize the maximum and minimum driver node networks for a given degree sequence? Taking a network produced by Algs.~\ref{alg:max} or \ref{alg:min} and measuring its degree correlations, however, would be misleading, since these algorithms provide only one out of many possible realizations of maximum or minimum driver node networks.
Instead, we investigate the range of possible realizations using link rewiring algorithms that preserve both the degree distribution of the network and the number of driver nodes.

We first map out the range of possible degree correlations for the maximum driver node realizations of the real networks listed in Table~1 of the supplementary material. For each network, we start by generating a bipartite representation of a maximum driver node network using Alg.~\ref{alg:max} with its corresponding node coloring. We then rewire the network to maximize the out-in degree correlation measured by the Pearson coefficient $r$, while preserving the degree distribution and the number of driver nodes. For this, we randomly select two links $(v^+,v^-)$ and $(w^+,w^-)$ and rewire them creating links $(w^+,v^-)$ and $(v^+,w^-)$ if (i)~they increase the out-in correlation, (ii)~the new links do not create double links, and (iii)~they do not violate the coloring, i.e., they do not connect two non-black nodes. The last condition ensures that the number of driver nodes does not change. We repeat this step until $r$ reaches a stationary value $r_\T{max}$. We then similarly find the minimum possible correlation $r_\T{min}$. And finally, we apply a very similar rewiring procedure to study the minimum driver node realization of the network produced by Alg.~\ref{alg:min}.

\begin{figure*}[ht!]
\centering
\includegraphics[scale=0.56]{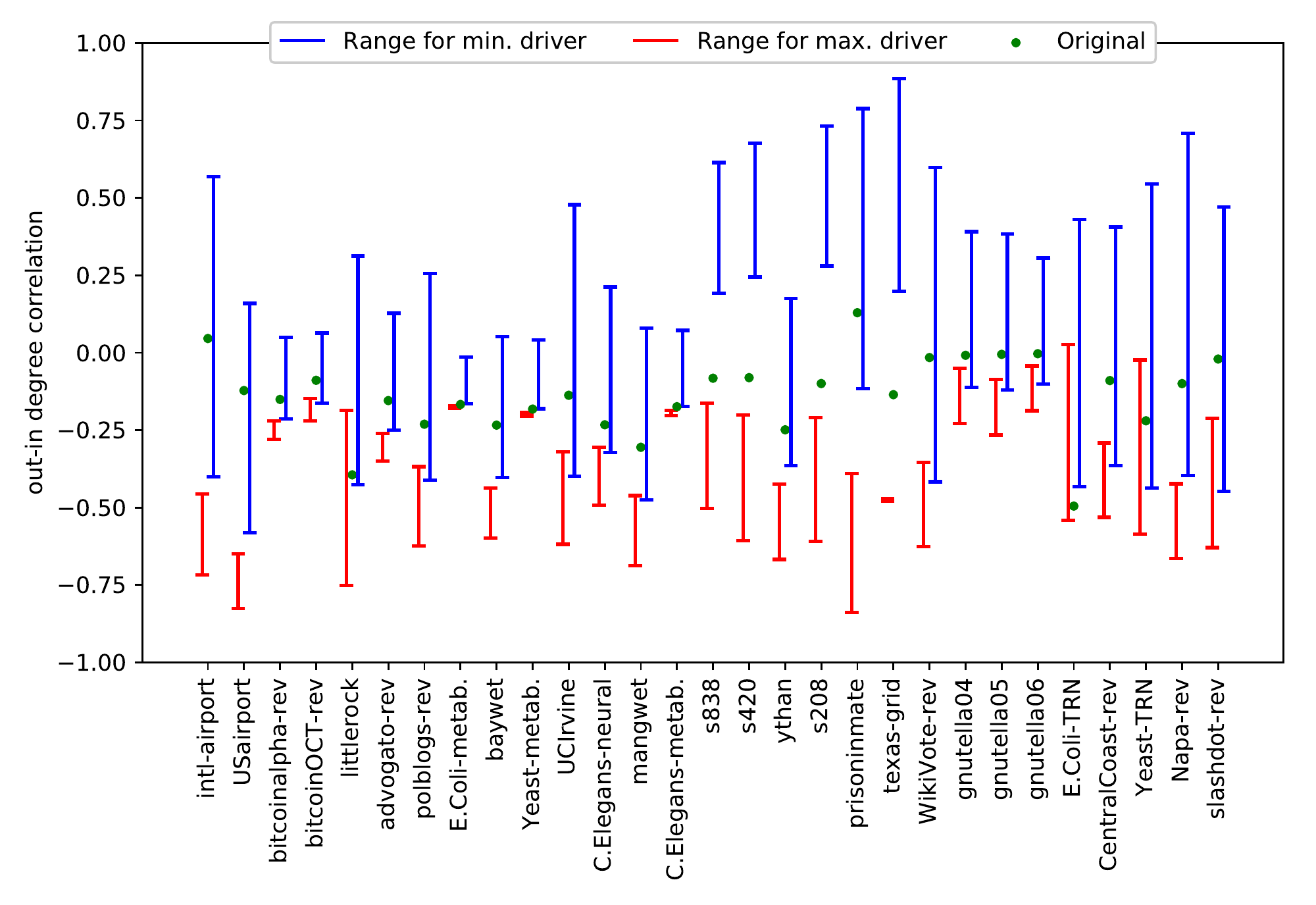}
  \caption{The range of out-in degree correlations for the maximum and minimum driver node realizations of real networks. The strength of the out-in degree correlation is measured by the Pearson correlation coefficient. The green dots provide the correlation values of the original networks, and the vertical lines indicate its possible values for the maximum and minimum driver node case accesible via edge rewiring while preserving the degree distribution and the number of required control signals.}\label{assortativity}
\end{figure*}

Figure~\ref{assortativity} shows the range of degree correlations of the maximum and minimum driver node realizations of real networks. Overall we find consistent results with Ref.~\cite{posfai2013effect}: maximum driver node networks are typically characterized by lower, while minimum driver node networks by higher out-in degree correlations. However, mapping out the range of possible correlations reveals that there is room for deviation from this pattern. For some networks, such as the Little Rock food web or the E. coli and yeast transcription networks, the possible correlation coefficient values for the maximum and minimum driver node realizations significantly overlap; therefore, some maximum driver node realizations have weaker or more negative in-out correlation than some minimum driver node realizations.
Our results show that using a simple correlation coefficient to summarize degree correlations in some cases may not be sufficient to predict the controllability of networks.

Note that our method to map out possible correlations is not exhaustive and may underestimate the range of coefficient values because (i)~the rewiring scheme itself is a heuristic and (ii)~we fix the coloring of the networks and other realizations with the same number of driver nodes may exist that do not correspond to that particular coloring.

\subsection{Control Profile}

\begin{figure}[ht!]
\hspace*{-1cm}
\includegraphics[scale=0.38]{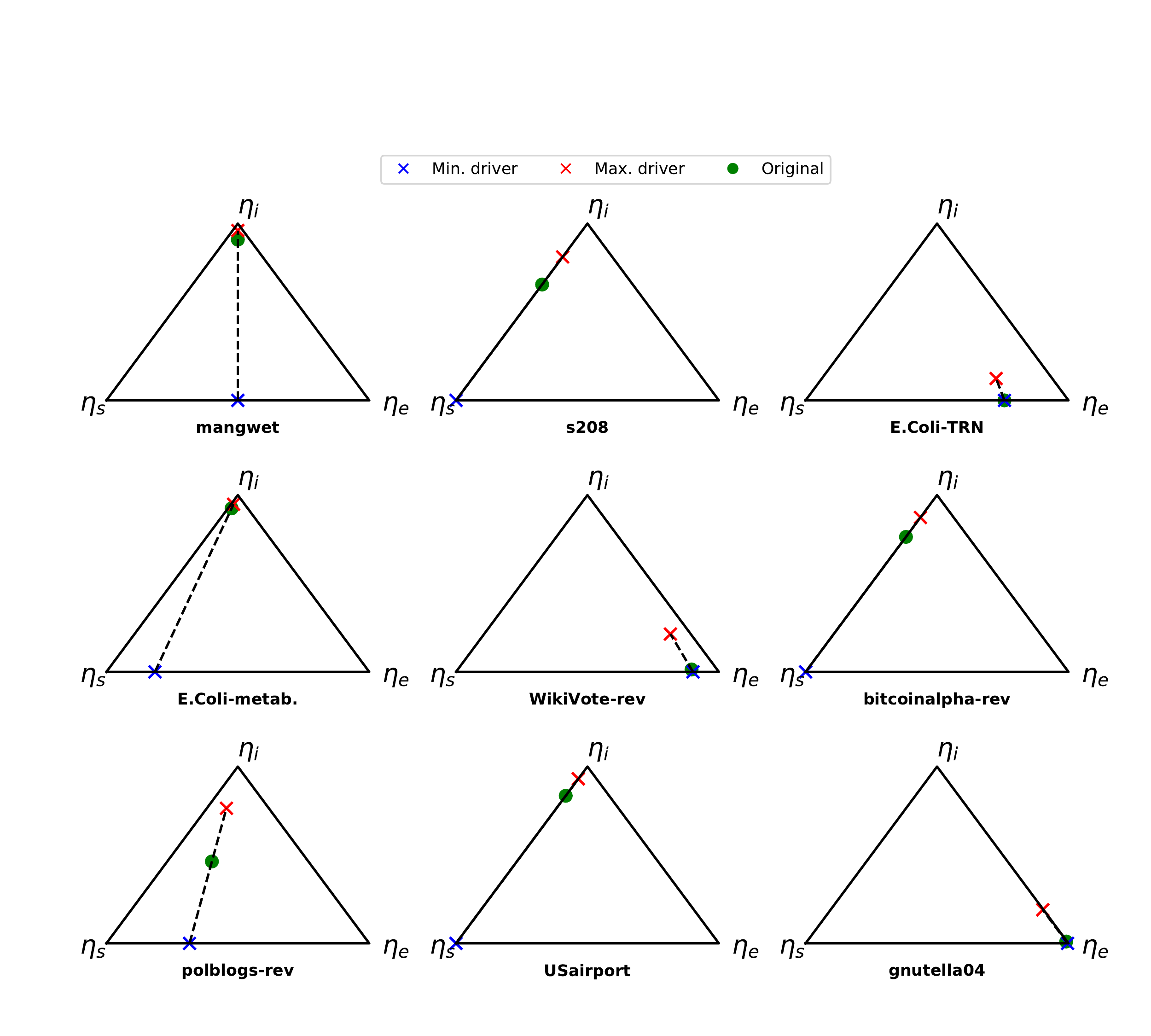}
  \caption{\label{fig:cntrlProfile} Control profiles of a selection of real networks. We show the control profile of the original network (green dot),  the maximum driver node realization (red cross), and the minimum driver node realization (blue cross). The dashed line indicates the region accessible by degree preserving rewiring.}
\end{figure}

The control profile of networks was introduced to characterize the origin of driver nodes~\cite{ruths2014control}. It classifies the driver nodes into three categories: (i)~Sources, drivers that correspond to nodes with zero in-degree; the number of sources is, therefore, $N_\T s = N_0^-$. (ii)~External dilations, drivers that are required if there is an excess of sinks compared to sources; the number of external dilations is $N_\T e=\max(0, N_0^+ - N_0^-)$. And (iii)~internal dilations, driver nodes that cannot be explained by the presence of sources or sinks, and are needed due to bottlenecks in the internal structure of the network; the number of internal dilations is $N_\T i = N_\T D - N_s - N_e$.
The control profile is then defined as $(\eta_\T s, \eta_\T e,\eta_\T i) = (N_\T s/N_\T D, N_\T e/N_\T D, N_\T i/N_\T D)$, where $\eta_\T s + \eta_\T e + \eta_\T i = 1$. In Ref.~\cite{campbell2015topological}, using degree preserved randomizations the authors demonstrated that the degree sequence of networks does not completely determine their control profile.
Here, using Algs.~\ref{alg:max} and \ref{alg:min} we precisely determine the range of the control profile where networks with given degree sequence are constrained to.

Figure~\ref{fig:cntrlProfile} shows the ternary plots of the control profiles of a selection of real networks. The control profile of a network is ultimately determined by three quantities: the number of sources $N_0^-$, the number of sinks $N_0^+$, and the number of driver nodes $N_\T D$. The number of sources and sinks is determined by the degree sequence; therefore, under degree preserving rewiring the control profile can only change through changes in $N_\T D$. This also means that the possible control profiles accessible through degree preserving rewiring are confined to a one dimensional segment (dashed line in Fig.~\ref{fig:cntrlProfile}).
One end of the segment corresponds to the minimum driver node realization of the network, indicated by a blue cross in Fig.~\ref{fig:cntrlProfile}. We showed that for all real and all reasonable model networks $N_\T D^\T{min} = \max(1, N_0^-,N_0^+)$; therefore the control profile can always reach a point where $\eta_\T i= 0$.
The other end of the segment corresponds to the maximum driver node realization of the network at 
\begin{equation}
\begin{split}
(\eta_\T s&, \eta_\T e,\eta_\T i) = \\
=& (N_\T s/N_\T D^\T{max}, N_\T e/N_\T D^\T{max}, 1-\max(N_0^-,N_0^+)/N_\T D^\T{max}) =\\
=&(N_\T s/N_\T D^\T{max}, N_\T e/N_\T D^\T{max}, 1-N_\T D^\T{min}/N_\T D^\T{max}),
\end{split}
\end{equation}
indicated by a red cross in Fig.~\ref{fig:cntrlProfile}.
Networks with $\eta_\T i\approx 1$ are called internal dilation dominated, and a network can be rewired to be internal dilation dominated if $N_\T D^\T{max}$ is much larger than $N_\T D^\T{min}$, for example, the \emph{mangwet} food web or the \emph{E. coli} metabolic network. Previously, it was demonstrated through random rewiring that the degree sequence of a network does not completely determine their control profile~\cite{campbell2015topological}. Here, we showed that our results allow us to exactly identify the range of profiles accessible through such rewiring.

\section{Conclusion}\label{sec:conclusion}

By relying on the concept of graphicality, our work introduces a novel set of tools for studying the controllability of complex networks.
We developed algorithms and analytical methods to investigate to what extent the degree sequence of a directed network constrains the number of driver nodes necessary to control the network.
We used these results to introduce \bl{structural control} complexity, a measure of how much more difficult it is to control a network beyond what is required given its degree sequence,
 and we applied our tools to study real and model networks. We showed, for example, that there exist networks that are characterized by a high number of driver nodes, yet have low \bl{structural control} complexity.
Furthermore, we demonstrated that our approach complements our existing knowledge and helps us better understand established results, such as the role of degree correlations in network controllability and the control profile of complex networks.
Future work may extend our approach to other notions of network control. For example, it was demonstrated that control through minimum driver node sets may have prohibitively high energy requirements~\cite{yan2012controlling,pasqualetti2014controllability,yan2015spectrum}. Therefore, it would be interesting to understand how degree sequence constrains such control energy. Also, this article focused on linear dynamics, future work should investigate scenarios where nonlinearity has to be taken into account, for example, when the goal of control is to switch between stable attractors~\cite{fiedler2013dynamics,zanudo2017structure}.

\ifCLASSOPTIONcompsoc
  \section*{Acknowledgments}
\else
  \section*{Acknowledgment}
\fi

We gratefully acknowledge support from the U.S. Army Research Office MURI Award No. W911NF-13-1-0340 and DARPA Award No. W911NF-17-1-0077. A. Ghasemi gratefully  acknowledges R. M. D'Souza and the department of computer science, University of California, Davis for their support during his visit.

\ifCLASSOPTIONcaptionsoff
  \newpage
\fi

\bibliographystyle{IEEEtran} 
\bibliography{refs}

\begin{IEEEbiography}[{\includegraphics[width=1in,height=1.25in,clip,keepaspectratio]{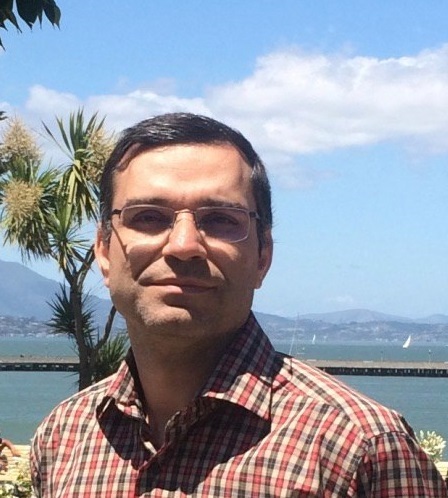}}]{Abdorasoul Ghasemi}
 received his Ph.D. degree from Amirkabir University of Technology (Tehran Polytechnique), Tehran, Iran in Electrical Engineering. He is currently an Associate Professor with the Faculty of Computer Engineering of K.N. Toosi University of Technology, Tehran, Iran. He has spent sabbatical leave with the department of computer science at the University of California, Davis, CA, the USA from April 2017 to August 2018. His research interests include network science and its engineering applications, communication networks, resource management in wireless networks, and applications of optimization and game theories in networking. 
\end{IEEEbiography}

\begin{IEEEbiography}[{\includegraphics[width=1in,height=1.25in,clip,keepaspectratio]{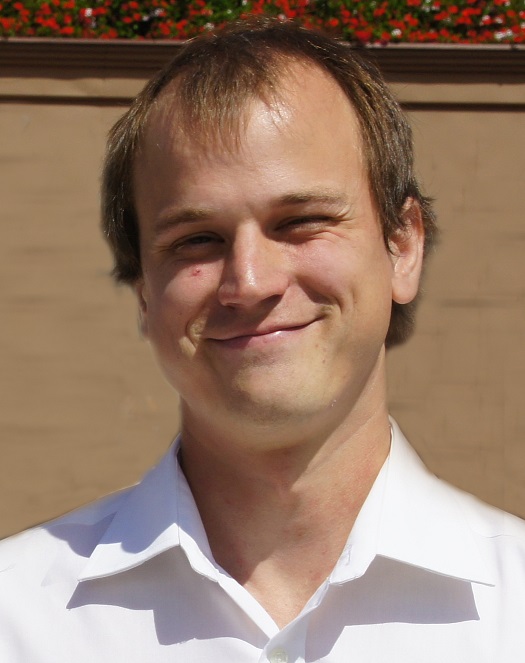}}]{M\'arton~P\'osfai}

received his PhD in statistical physics at the Eötvös University, Budapest under the supervision of Prof. Gábor Vattay. During his PhD he spent two years at Northeastern University, Boston and six months at TU Berlin. Currently, he is a postdoctoral researcher at the University of California, Davis working together with Prof. Raissa D’Souza. His research interests span various topics of network science, including structure of complex networks, modeling the emergence of social order and stability, and using methods of control theory to understand the organization of complex networks.
\end{IEEEbiography}

\begin{IEEEbiography}[{\includegraphics[width=1in,height=1.25in,clip,keepaspectratio]{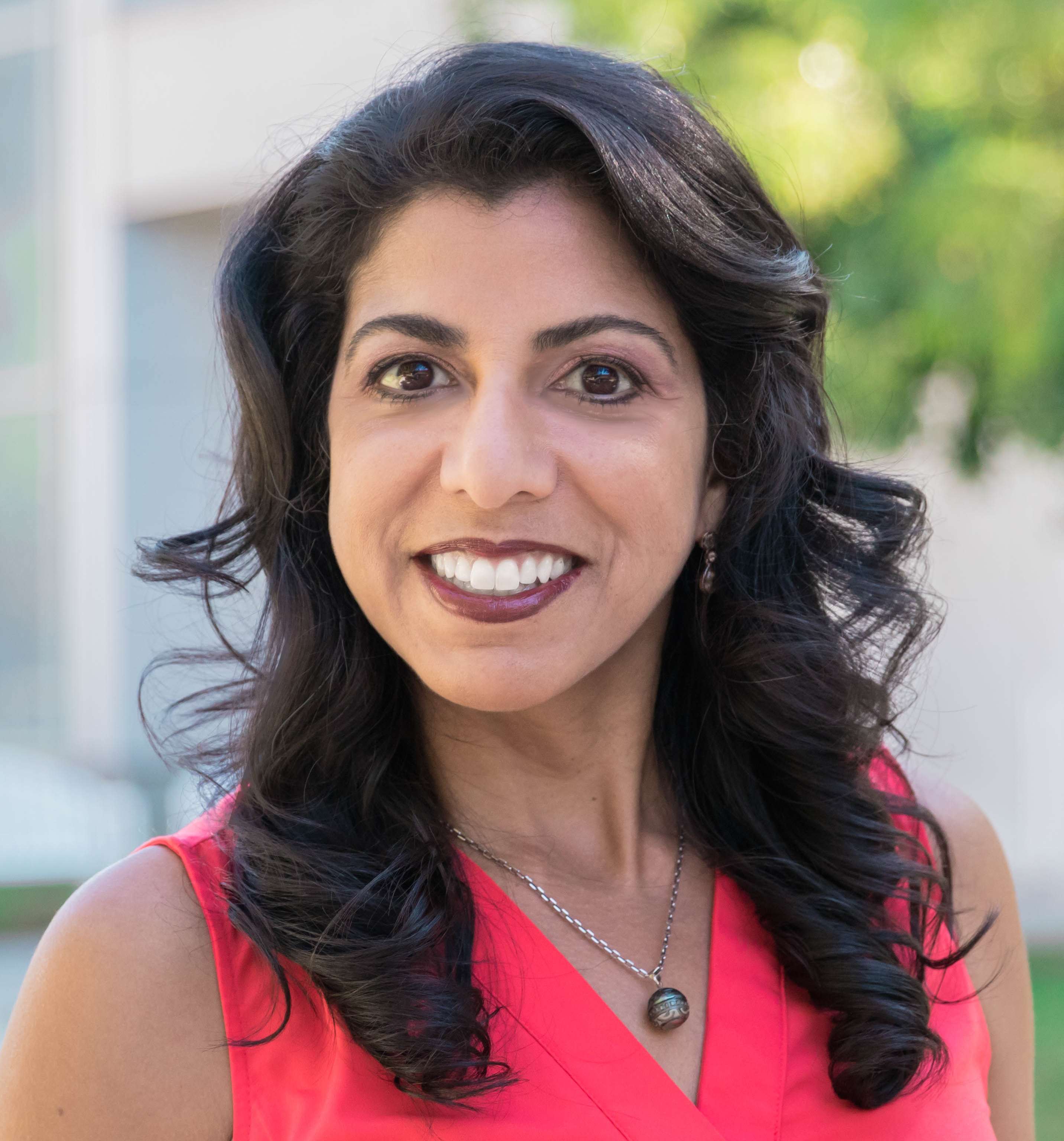}}]{Raissa M. D'Souza}
is Professor of Computer Science and of Mechanical Engineering at the University of California, Davis, as well as an External Professor at the Santa Fe Institute. She received a PhD in Statistical Physics from MIT in 1999, then was a postdoctoral fellow, first in Fundamental Mathematics and Theoretical Physics at Bell Laboratories, and then in the Theory Group at Microsoft Research. Her interdisciplinary work on network theory and complex systems spans the fields of statistical physics, theoretical computer science and applied math, and has appeared in journals such as Science, PNAS, and Physical Review Letters. She is a Fellow of the American Physical Society (class of 2016), a Fellow of the Network Science Society (class of 2019), and has received numerous honors such as the inaugural Euler Award of the Network Science Society, the 2018 ACM Test-of-Time award, and the 2017 UC Davis College of Engineering Outstanding Mid-Career Faculty Research Award. She serves on the editorial board of numerous international mathematics and physics journals, has organized key scientific meetings like NetSci 2014, and was a member of the World Economic Forum's Global Agenda Council on Complex Systems. She served as President of the Network Science Society, 2015-18, and as Outgoing President 2018-19.
\end{IEEEbiography}



\end{document}


\title{\LARGE{Supplementary material}\\ \vspace{1mm}\LARGE{Diversity of structural controllability of complex networks with
given degree sequence}}

\author{Abdorasoul~Ghasemi,
        M\'arton~P\'osfai, and~Raissa M. D'Souza
}

\markboth{IEEE Transactions on Network Science and Engineering}%
{Shell \MakeLowercase{\textit{et al.}}: Bare Demo of IEEEtran.cls for Computer Society Journals}

\maketitle

\section{Model Networks: Derivations and approximations}
In this section, we provide more details about the modified static model with a controlled fraction
of unconnected nodes. We use this model to systematically investigate the effects of the fraction of sources and sinks on $n_\T D$, $n_\T D^\T{min}$, and $n_\T D^\T{max}$.  

We start by discussing how well the binomial distribution approximates the degree distribution of the static model or its modified version. Note that we use the binomial approximation in Eq. (8) to respectively calculate the fraction of non-black and zero-degree nodes in Sec.~4 of the main text to numerically find the lower and upper bounds for the modified static model.  Next, we discuss the tail of the degree distribution of the generated networks with the modified static model using the approach of Ref. \cite{catanzaro2005analytic}, and we show that it has power-law tail property. Finally, we provide more details about the derivations and asymptotic behavior of lower- and upper-bounds.

\subsection{Degree distribution of modified static model}
Recall the modified static model in which we assign a weight $w_i=i^{-\alpha}/(\sum_j j^{-\alpha})$ to nodes $i=1,2,\ldots,N-N_0$ where $\alpha \in [0,1)$, and $w_i=0$ to nodes $i=(N-N_0+1),\ldots,N$. To generate a network, we start with $N$ unconnected nodes, and we randomly select two nodes $i$ and $j$ with probability $w_i$ and $w_j$, respectively, and if there is no directed link from node $i$ to $j$, we connect them. We repeat this step until $L$ links are added. 

Take node $i$, if we assume that the consecutive trials of adding links are independent, i.e., we allow multiple links between a pair, the out-degree (in-degree) of node $i$  follows a binomial distribution given by $\binom{L}{k} w_i^k (1~-~w_i)^{L-k}$. Therefore, the fraction of nodes with out-degree (in-degree) $k$ is given by 

\begin{equation} \label{equ: binomial_approx}
p(k) = \frac{1}{N} \sum_{i=1}^{N} \binom{L}{k} w_i^k (1 - w_i)^{L-k}.
\end{equation}

However, since multiple links between two nodes are not allowed, the consecutive additions of links are not independent, and the binomial distribution provides an approximation. In the following we calculate the expected total number of multi-links in the generated networks and we show that for networks with $\gamma > 3$ the expected number of multi-links per node, i.e., the density of muti-links \cite{newman2018networks}, vanishes as $\frac{1}{N}$ where $N$ is the number of nodes, and hence its impact on the degree distribution is negligible. For $2 < \gamma \leq 3$, the expected density of multi-links vanishes more slowly than $\frac{1}{N}$ as the network size is increased and affects the degree distribution of the networks making Eq.~(\ref{equ: binomial_approx}) an approximation. Nevertheless, the density of multi-links is not uniform across different degrees and affects the high degree nodes more strongly, making the approximation error higher for large $k$ whereas the density of multi-links between low degree nodes is negligible. Note that we, respectively, need to numerically calculate the fraction of low degree non-black nodes and the fraction of zero in- and out-degree nodes in Sec.~4 of the main text to compute the upper and lower bounds. Therefore, the binomial approximation works well for computing both bounds.

The probability of adding a specific directed link from $i$ to $j$ is $p_{ij} = w_i w_j$. Let $X_{ij}$ denote the random variable corresponding to the number of directed links between $i$ and $j$. The probability of having multi-links, i.e., at least two directed links between $i$ and $j$ is given by
\begin{equation}\label{eq:multiple_links}
\Pr[X_{ij} \geq 2] = 1 - \Pr[X_{ij}=0] - \Pr[X_{ij}=1] = 1 - \left(1-p_{ij}\right)^L - L p_{ij} \left(1 - p_{ij}\right)^{L-1}. 
\end{equation}
Summing this probability over all node pairs, we can compute the expected total number of multiple links as
\begin{equation}
 M = \sum_{i,j=1}^{N}Pr[X_{ij} \geq 2]. 
\end{equation}

\subsubsection{Exponent $2 \alpha < 1$}

For $2 \alpha < 1$ (corresponding to $\gamma > 3$), we can simplify (\ref{eq:multiple_links}) to find an upper bound for $M$ using $(1-p_{ij})^L \approx 1-Lp_{ij}$ for 
$p_{ij}=w_iw_j < 1$ and $L p_{ij} \ll 1$. Since $w_i$ gets its maximum value for $i=1$, we need to have
\begin{equation}\label{equ: multi_approx}
L \frac{i^{-\alpha} j^{-\alpha}}{\left[\sum_{k=1}^{N-N_0} k^{-\alpha}\right]^2} \leq L \frac{(1-\alpha)^2}{\left[\left(N-N_0+1\right)^{1-\alpha}-1\right]^2} \approx \frac{c}{1-n_0} \times \frac{(1-\alpha)^2}{(N-N_0)^{1-2\alpha}} \ll 1,
\end{equation}
which holds if $2 \alpha < 1$ as $N-N_0$ increases. In (\ref{equ: multi_approx}), $c=\frac{L}{N}$ is the average degree of network and we use the fact that for decreasing function $j^{-\alpha}$
\begin{equation}\label{eq:upper_bound}
\frac{\left(N+1\right)^{1-\alpha} - 1}{1-\alpha} = \int_{1}^{N+1}j^{-\alpha}dj \leq \sum_{j=1}^{N}j^{-\alpha} \leq \int_{0} ^{N} j^{-\alpha}\,dj = \frac{N^{1-\alpha}}{1-\alpha},  0 < \alpha < 1.
\end{equation}
Therefore, in this regime we can use the approximation $\Pr[X_{ij} \geq 2] \approx L(L-1)p_{ij}^2$. We have 
\begin{equation}\label{eq:approx_multilinks}
\sum_{i,j=1}^{N}p_{ij}^2 = \sum_{i,j=1}^{N-N_0} w_i^2 w_j^2 = \left[ \sum_{i=1}^{N-N_0} w_i^2\right]^2 = \left[ \sum_{i=1}^{N-N_0} {\frac{i^{-2\alpha}}
{\left(\sum_{j=1}^{N-N0}j^{-\alpha}\right)^2}}\right]^2 \leq \frac{\left(1-\alpha\right)^4}{\left[\left(N-N_0+1\right)^{1-\alpha} - 1\right]^4} \left[ \sum_{i=1}^{N-N_0} {i^{-2\alpha}}\right]^2.
\end{equation}
By applying  the right hand side of (\ref{eq:upper_bound}) in (\ref{eq:approx_multilinks}) for $2 \alpha < 1$ we have
\begin{equation}\label{eq:gamma3}
M \approx  L(L-1) \sum_{i,j=1}^{N}p^2_{ij}  \leq \frac{L(L-1)\left(1-\alpha\right)^4}{{\left[\left(N-N_0+1\right)^{1-\alpha} - 1\right]^4} } \frac{\left(N-N_0\right)^{2-4\alpha}}{\left(1-2\alpha\right)^2} \approx \frac{\left(1-\alpha\right)^4}{\left(1-2\alpha\right)^2} \left(\frac{L}{N-N_0}\right)^2 \approx \frac{\left(1-\alpha\right)^4}{\left(1-2\alpha\right)^2} \left(\frac{c}{1-n_0}\right)^2.
\end{equation}
 That is, for $\gamma > 3$,  $M$ is bounded by a constant and is independent of $N$. Therefore, the density of multi-links, $\frac{M}{N}$, vanishes as $N-N_0$ increases; hence, its impact on binomial approximated degree distribution is negligible as it is shown in Fig.~\ref{fig: supp_Fig1}(a).


\subsubsection{Exponent $2\alpha>1$}

\begin{figure}[h!]

\centering
   \begin{subfigure}{0.49\textwidth}
   \includegraphics[width=\linewidth]{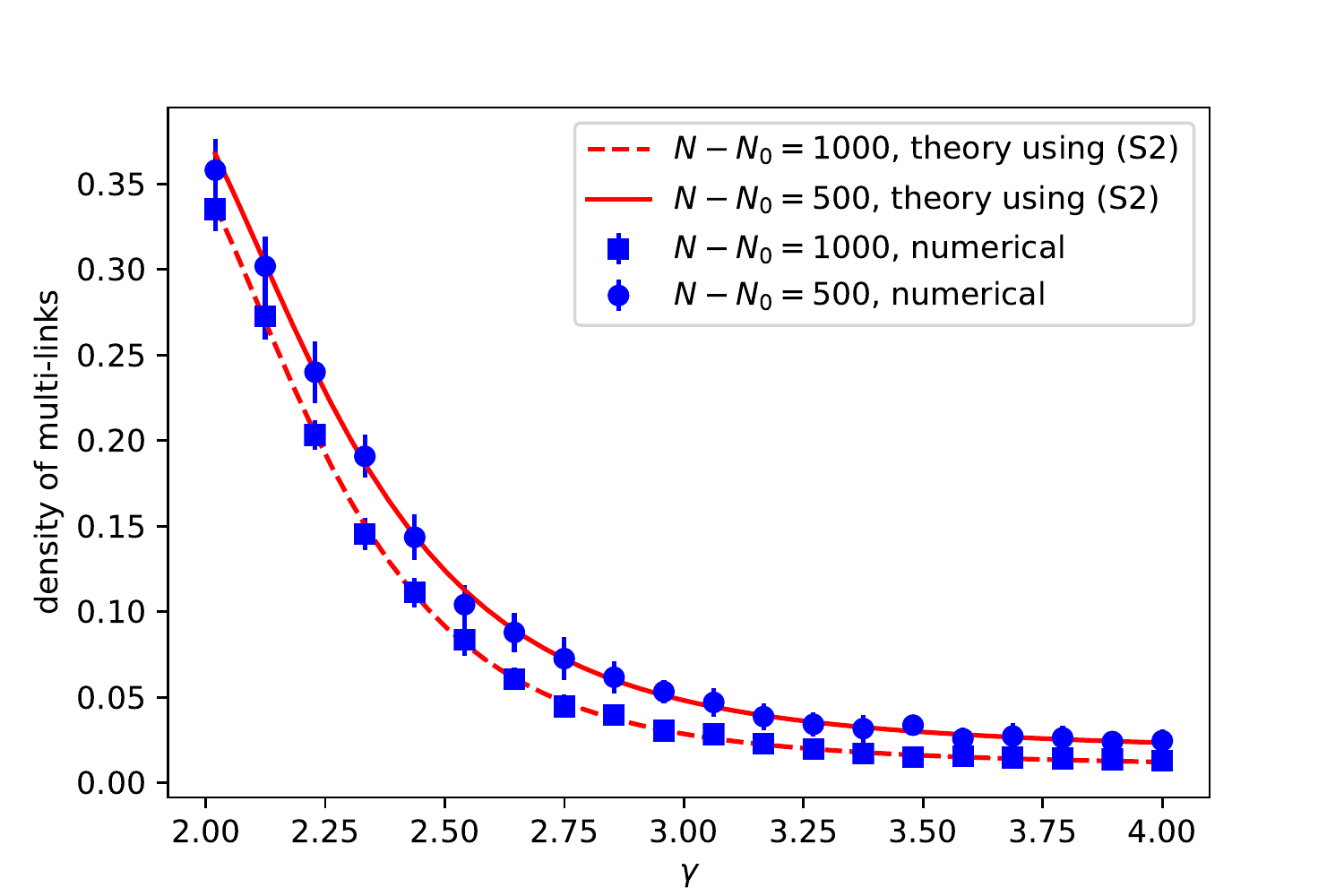}
\end{subfigure}
\begin{subfigure}{0.49\textwidth}
   \includegraphics[width=\linewidth]{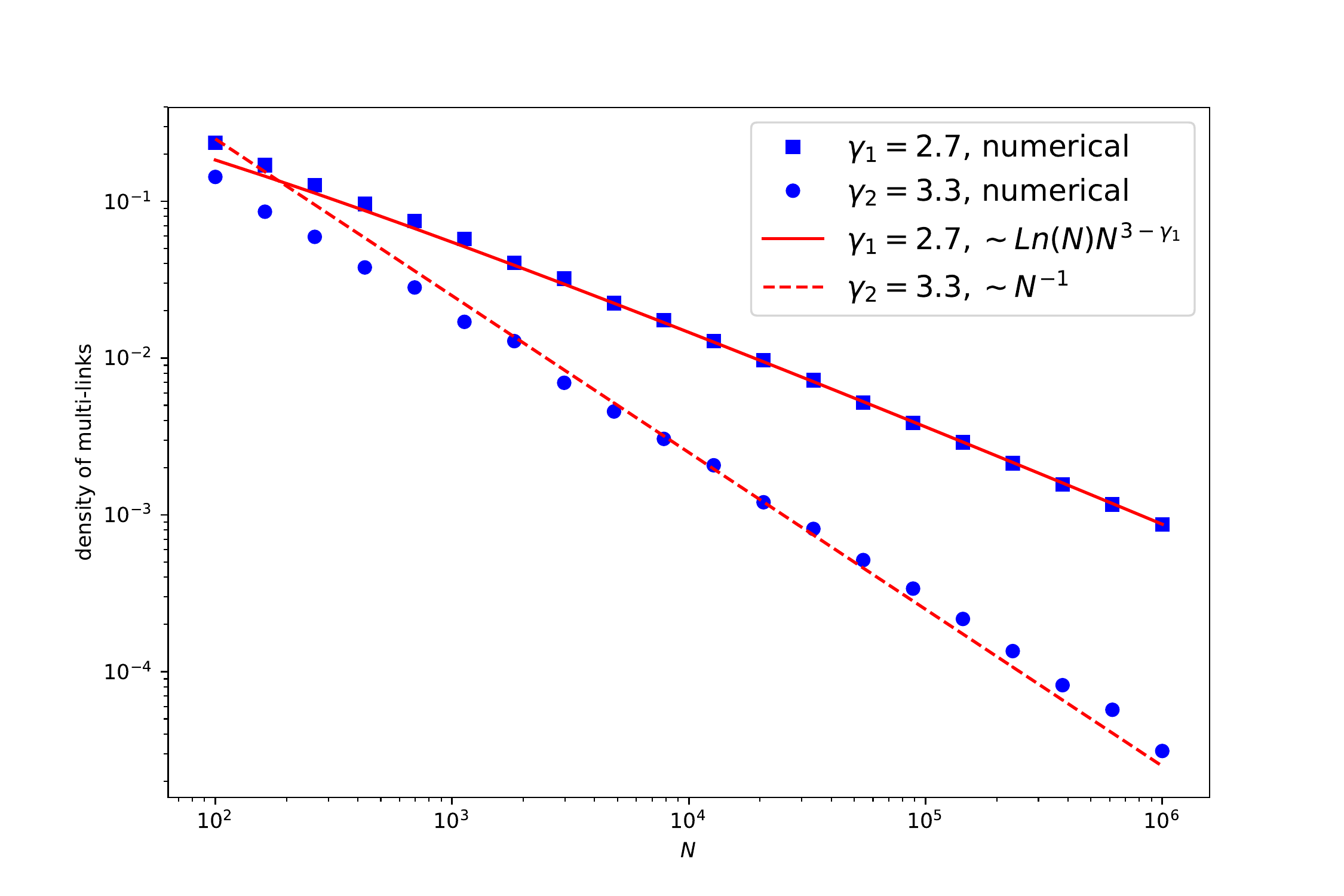}
\end{subfigure}

\caption{ \label{fig: supp_Fig1} (a) The density of multi-links against $\gamma$ for networks with $N-N_0=1000, 500$ nodes and $c=4$. Numerical results are the average of 20 independent realizations for each scenario, and error bars indicate the standard deviation. (b) The density of multi-links against $N$ for $c=4$, $\gamma_1=2.7$, and $\gamma_2=3.3$. The density of multi-links for $\gamma_2 > 3$ vanishes  as $\frac{1}{N}$ and the total expected number of links is bounded by a constant independent of $N$. The density of multi-links for $\gamma_1 < 3$ vanishes more slowly than $\frac{1}{N}$. Numerical results are the average of 20 independent realizations for each $N$.}
\end{figure}

If the exponent $\alpha$ satisfies $2\alpha>1$ (corresponding to $2 < \gamma \leq 3$), the expected number of links between some pairs of nodes will diverge with network size as $(N-N_0)^\beta$, where $\beta>0$. For example, the expected number of links between node $i=1$ and $j=2$ is
\begin{equation}
Lp_{12} = L \frac{2^{-\alpha}}{\left[\sum_{k=1}^{N-N_0} k^{-\alpha}\right]^2} \approx \frac{2^{-\alpha} c (1-\alpha)^2 }{1-n_0} (N-N_0)^{2\alpha-1}.
\end{equation}
Using the approximation $(1-p_{ij})^L\approx\Exp(-Lp_{ij})$ we see that the probability that no or only a single link exists between such node pairs decays exponentially; therefore counting the pairs of nodes for which $Lp_{ij}>1$ allows us to estimate the density of multi-links in leading order of $N$. The condition for this can be written as
\begin{align}
1 &< cN\frac{(ij)^{-\alpha}}{\sum_{k=1}^{N-N_0}k^{\alpha}}\approx \frac{c(1-\alpha)^2}{1-n_0}\frac{(ij)^{-\alpha}}{(N-N_0)^{1-2\alpha}} \\
j &<  \frac{1}{i} \left(\frac{(1-n_0) (N-N_0)^{1-2\alpha}}{c(1-\alpha)^2}\right)^{-1/\alpha}=\frac{A}{i},
\end{align}
where we introduced $A$ to simplify equations.
Therefore, the number of multi-links is
\begin{equation}
M = \sum_{Lp_{ij}>1}1\approx \int_1^{N-N_0}di\int_{1}^{A/i} dj \sim \ln(N-N_0)(N-N_0)^{1-\frac{1}{\alpha}}=\ln(N-N_0)(N-N_0)^{3-\gamma},
\end{equation}
this means that the density of multi-links for this case also decays to zero as the network size is increased, albeit slower than for the $2\alpha<1$ case. See Fig.~\ref{fig: supp_Fig1}(b). The decay is particularly slow, for degree exponents close to $2$, indeed we find the most discrepancy between theory for $\gamma\approx 2$, see Figs.~4 and 5 in the main text.

\begin{figure}[ht!]
\centering
\includegraphics[scale=0.45]{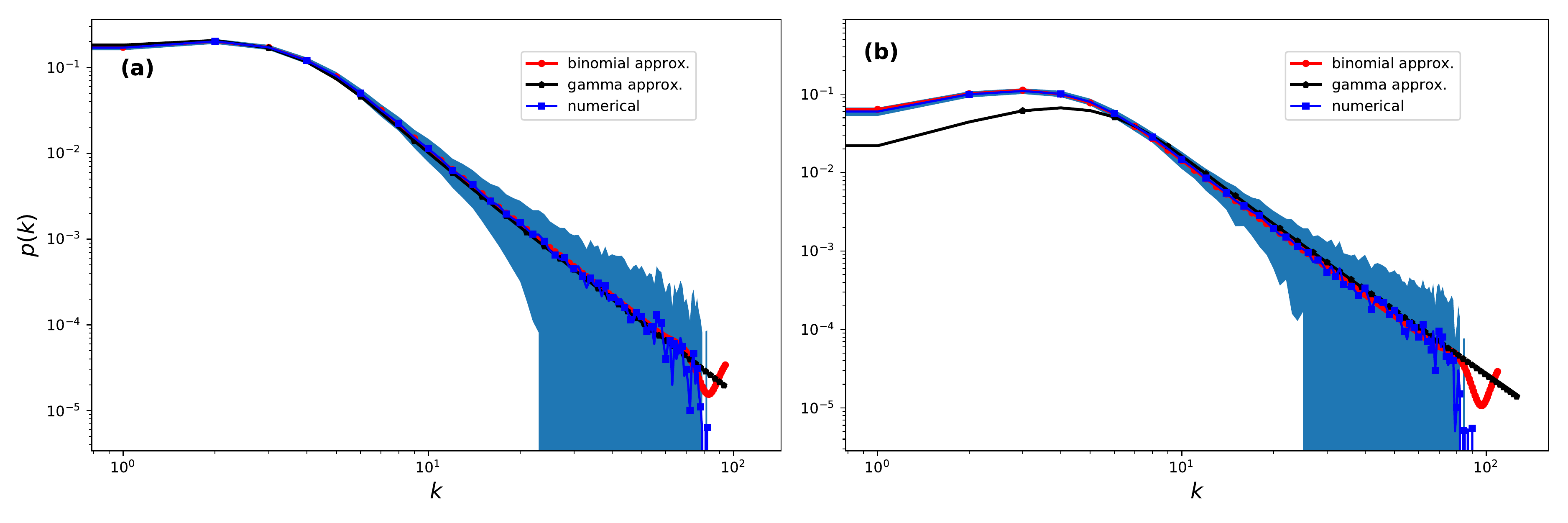}
  \caption{\label{fig:supp_degree} The degree distribution of empirical constructed networks, approximated by the binomial distribution with $\gamma=2.7, c=4$, and (a) {$N-N_0=1000$} nodes and (b) $N-N_0~=~500$ nodes. The gamma distribution approximates the tail of the distribution. Numerical results are the average of 200 independent networks, and shaded areas show the standard deviation.}
\end{figure}

Therefore, for $2 < \gamma \leq 3$ the number of multiple links and hence the induced error in binomial approximated degree distribution is increased.  However, in this regime, the high degree nodes appear, and the density of multi-links is higher for these nodes. That is, the occurrence of multi-links is more probable between high degree nodes and the probability of multi-links between low degree nodes is negligible, as it is shown in Fig. \ref{fig:supp_degree}. Therefore, the binomial distribution provides a good approximation for the fraction of low degree nodes as we need to find the lower and upper bounds. Recall that the maximum degree of non-black nodes $k_0$ is constant and independent of $N$ in finding the upper bound, and to find the lower bound, we need the fraction of zero degree nodes.

\newpage
\subsubsection{Tail of degree distribution}
Next, we show that as in the static model, the tail of the modified static model also follows the power-law distribution. We start using the fact that the binomial distribution can be approximated by Poisson distribution for sufficiently large values of $L$ as

\begin{equation} \label{equ: poisson_approx}
\binom{L}{k} w_i^k (1 - w_i)^{L-k} \approx \frac{\lambda_i^k e^{-\lambda_i}}{k!}, \lambda_i=L\,w_i.
\end{equation}

Note that using $w_i \approx \frac{(1-\alpha)}{N-N_0}\left(\frac{i}{N-N_0}\right)^{-\alpha}$, we have $\lambda_i=L\,w_i \approx \frac{c(1-\alpha)}{1-n_0}\left(\frac{i}{N-N_0}\right)^{-\alpha}$ which remains constant as $L$ increases; making (\ref{equ: poisson_approx}) a reasonable approximation. We have:
\begin{flalign}\label{equ: gamma_approx}
p(k) &\approx n_0\delta_{0,k} + \frac{1}{N} \int_{0}^{N-N_0}  \frac{e^{-\lambda_i}{\lambda_i}^k}{k!}\dif i \approx n_0\delta_{0,k} + \frac{1}{N} \int_{\infty}^{\frac{c(1-\alpha)}{1-n_0}} \frac{e^{-\lambda_i}{\lambda_i}^k}{k!} \left[\frac{1-n_0}{c(1-\alpha)}\right]^{\frac{-1}{\alpha}}\left(N-N_0\right)\left(\frac{-1}{\alpha}\right)\lambda_i^{\frac{-1}{\alpha}-1} \dif \lambda_i = &&\\ \nonumber
&= n_0\delta_{0,k} + (1-n_0) \frac{ \left(c[1-\alpha]/[1-n_0]\right)^{1/\alpha}}{\alpha \Gamma(k+1)} \int_{\frac{c(1-\alpha)}{1-n_0}}^{\infty} {e^{-\lambda_i}{\lambda_i}^{k-\frac{1}{\alpha}-1}} \dif \lambda_i = &&\\ \nonumber
&= n_0\delta_{0,k} + (1-n_0) \frac{ \left(c[1-\alpha]/[1-n_0]\right)^{1/\alpha}}{\alpha} \times \frac{\Gamma(k-1/\alpha,c[1-\alpha]/[1-n_0])}{\Gamma(k+1)},
\end{flalign}
where $\Gamma(a,x)$ is upper incomplete gamma function and is given by $\Gamma(a,x) = \int_{x}^{\infty}e^{-t} t^{a - 1} \dif t$. 

Since $\Gamma(a,x) \to \Gamma(a)$ for $a \to \infty$, we find that the asymptotic behavior of the degree distribution for large $k$ is:
\begin{equation}
p(k) \approx (1-n_0) \frac{ \left(c[1-\alpha]/[1-n_0]\right)^{1/\alpha}}{\alpha} \times \frac{\Gamma(k-1/\alpha)}{\Gamma(k+1)} \propto k ^{-(1+\frac{1}{\alpha})},
\end{equation}
where we use $\frac{\Gamma(k+a)}{\Gamma(k+b)} \propto k^{a-b} \left(1 + O(\frac{1}{k})\right)$ for $k \to \infty$. Therefore, the tail of the degree distribution of the modified static model has power-law property. See Fig. \ref{fig:supp_degree}.

\subsection{Lower bound}

The lower bound $n^\T{LB}_\T D$ is equal to the fraction of zero-degree nodes, from equation (\ref{equ: gamma_approx}) we get 
\begin{align}
\begin{split}\label{eq:LB_anal}
n^\T{LB}_\T D &= p(0) \approx n_0 + (1-n_0)\frac{(c[1-\alpha]/[1-n_0])^{1/\alpha}}{\alpha}\times \Gamma(-1/\alpha,c[1-\alpha]/[1-n_0]).
\end{split}
\end{align}

We can use the asymptotic behavior of $\Gamma(a,x)$ to find the asymptotic behavior of $n^\T{LB}_\T D$ in different scenarios. Specifically, we have \cite{jameson2016incomplete}: 

\begin{equation}
\Gamma(a,x) = x^{a-1}e^{-x}\left(1+O(1/x)\right) {\rm as}\, x\to \,\, \infty.
\end{equation}

Therefore for large values of $c$, the asymptotic behavior of the $n^\T{LB}_\T D$ is given by:
\begin{align}
n^\T{LB}_\T D -n_0 = (1-n_0)\frac{(Ac)^{1/\alpha}}{\alpha}\Gamma(-1/\alpha,Ac) \approx (1-n_0)\frac{(Ac)^{1/\alpha}}{\alpha}\left(Ac\right)^{-1/\alpha - 1} e^{-Ac} = (1-n_0)\frac{e^{-Ac}}{\alpha Ac}, 
\end{align}
where $A=(1-\alpha)/(1-n_0)$. That is, $n^\T{LB}_\T D \to n_0 $ for large values of $c$ as we see in Fig.~4~(b) of the main text. 

To investigate the asymptotic behavior of $n^\T{LB}_\T D$ as $\gamma \to 2^+$ (corresponding to $\alpha \to 1^-$), we use the fact that:

\begin{equation}
\Gamma(a,x) \approx \frac{-x^a}{a}, a < 0 \,\,\,\,{\rm as} \,\, x \to 0^+.
\end{equation}

Let $B = c/(1-n_0)$, we have:
\begin{equation}
n^\T{LB}_\T D -n_0 \approx (1-n_0) \frac{\left(B[1-\alpha]\right)^{1/\alpha}}{\alpha} \Gamma(-1/\alpha, B[1-\alpha]) \approx (1-n_0) \frac{\left(B[1-\alpha]\right)^{1/\alpha}}{\alpha} \times \frac{-\left(B[1-\alpha]\right)^{-1/\alpha}}{-1/\alpha} = 1-n_0.
\end{equation}
 Therefore, $n^\T{LB}_\T D \to 1$ as $\gamma \to 2$.

Similarly, $n^\T{LB}_\T D \to n_0$ for large values of $\gamma$ (corresponding to $\alpha \to 0^+$). The reason is that we have \cite{jameson2016incomplete}:
\begin{equation}
\Gamma(a,x) \leq e^{-x} \frac{x^a}{-a},  \,\, a < 0.
\end{equation}
Therefore, the upper bound of $n^\T{LB}_\T D$ is given by:
\begin{align}
n^\T{LB}_\T D -n_0 \approx (1-n_0) \frac{\left(B[1-\alpha]\right)^{1/\alpha}}{\alpha} \Gamma(-1/\alpha, B[1-\alpha]) & \leq  (1-n_0) \frac{\left(B[1-\alpha]\right)^{1/\alpha}}{\alpha} \frac{\left(B[1-\alpha]\right)^{-1/\alpha}}{1/\alpha} e^{-B[1-\alpha]} \\ \nonumber &= (1-n_0)e^{-B[1-\alpha]},
\end{align}
where for small values of $\alpha$ and sufficiently large values of $B$ we have $n^\T{LB}_\T D \to n_0$. 

\subsection{Upper bound}

To find the closed form solution and the asymptotic  behavior of $n^\T{UB}_\T D$, recall that we color all nodes black with degree larger than some degree $k_0$ and a $q$ fraction of nodes that have degree $k_0$ and we require that at least half of the links must be adjacent to black nodes.  Let $A = (1-n_0) \frac{ \left(c[1-\alpha]/[1-n_0]\right)^{1/\alpha}}{\alpha}$ in (\ref{equ: gamma_approx}), $k_0$ is given by:

\begin{equation}\label{equ: k0}
\int_{k_0} ^{\infty} A k^{-\gamma+1}\dif k = 0.5c = 0.5 \int_{1}^{\infty} A k ^{-\gamma+1} \dif k.
\end{equation}

Solving (\ref{equ: k0}), we have $k_0 = 2 ^ {\frac{-1}{\gamma-2}}$. Next, the fraction of black nodes is given by:
\begin{equation}
n_{\T B} \approx \int_{k_0} ^{\infty} p_k \dif k = \frac{A}{\gamma-1} k_0^{\gamma-1} = (1-n_0) 2 ^{-\frac{\gamma-1}{\gamma-2}},
\end{equation}
where $A = (\gamma-1) (1-n_0)$ noting that we have $n_0 + \int_{1}^{\infty}A k^{-\gamma} \dif k = 1$.  Therefore, as $\gamma$ approaches 2 a vanishing fraction of nodes will be colored black and $n^\T{UB}_\T D$ approaches 1. 
 
\clearpage

\section{Data sets}

The descriptions of the datasets and the results for the minimum (lower bound) and maximum (upper bound) number of driver nodes for each network are summarized in Table~\ref{tab:realnets}.

\begin{table}[h]
\caption{\label{tab:realnets}Real networks and their properties. $N, N^-_0, N^+_0$: number of nodes, sources, and sinks, $L$: number of links, $N_\T D , N_\T D^\T{rand}(\T{std})$:  number of drivers for original network and its average (standard deviation) for random rewired networks, $N_\T D^\T{min} (N_\T D^\T{LB})$: min. number of driver (lower bound), $N_\T D^\T{max} (N_\T D^\T{UB})$: max. number of driver (upper bound). } 
\centering
\resizebox{\textwidth}{!}{%
\begin{threeparttable}

\begin{tabular}{r|llllllllll}

\hline 
 Data Set & Network & $N, N^-_0, N^+_0$ & $L$ &$N_\T D$& $N_\T D^\T{rand}(\T{std})$& $N_\T D^\T{min} (N_\T D^\T{LB})$& $N_\T D^\T{max} (N_\T D^\T{UB})$& Ref.\\
\hline\hline 
Food Web & mangwet & 97, 1, 2 & 1492 & 22 & 3.9 (1.2) & 2 (2) & 50 (54) & \cite{baird1998assessment} \\
 & baywet & 128, 1, 2 & 2106 & 30 & 7.5 (1.7) & 2 (2) & 69 (70) & \cite{ulanowicz2005network} \\
 & littlerock & 183, 1, 1 & 2476 & 99 & 46.6 (3.5) & 1 (1) & 120 (123) & \cite{martinez1991artifacts} \\
 & ythan & 135, 52, 1 & 597 & 69 & 61.4 (1.9) & 52 (52) & 100 (100) & \cite{dunne2002food} \\
\hline 
Electric Circuit & s208 & 122, 10, 1 & 189 & 29 & 24.1 (2.7) & 10 (10) & 53 (53) & \cite{milo2002network}\\
& s420 & 252, 18, 1 & 399 & 59 & 49.1 (3.8)& 18 (18) & 110 (110) &  \cite{milo2002network}\\
& s838 & 512, 34, 1 & 819 & 119 & 100.5 (4.5)& 34 (34) & 224 (224) & \cite{milo2002network}\\
\hline
Transcription & E. Coli-TRN & 418, 76, 312 & 519 & 314 & 315.3 (1.6) & 312 (312) & 356 (356) & \cite{milo2002network} \\
& Yeast-TRN & 688, 96, 557 & 1079& 565 & 558.2 (1.0) & 557 (557) & 605 (605) & \cite{milo2002network}\\
\hline
Metabolic & C.Elegans-metab. & 1173, 40, 37 & 2864 & 354 & 236.4 (6.8) & 40 (40) & 599 (599) & \cite{jeong2000large}\\
& E.Coli-metab. & 2275, 53, 65 & 5763 & 870 & 505.1 (11.1) & 65 (65) & 1262 (1262) & \cite{jeong2000large}\\
& Yeast-metab. & 1511, 43, 40 & 3833 & 497 & 316.5 (8.6) & 43 (43) & 825 (825) & \cite{jeong2000large}\\\hline
Social & Prison inmate & 67, 4, 7 & 182 & 9 & 9.7 (1.2) & 7 (7) & 30 (30) & \cite{van2003evolution}\\

 & UCIrvine & 1899, 37, 549 & 20296 & 614 & 610.8 (6.7) & 549 (549) & 1525 (1525) & \cite{opsahl2009clustering}\\
 
 & slashdot-friends-rev\tnote{1} & 11227, 10559, 9 & 30914 & 10577 & 10563.9(2.2) & 10559 (10559) & 10942 (10942) & \cite{tang2012inferring}\\
 
 \hline
\makecell{Web of Trust \\(Social Influence)}   

& Central Coast\tnote{2} &  943, 718, 152 & 1127 & 721 & 719.4 (0.86) & 718 (718) & 792 (792) & \cite{levy2018innovation} \\

& Napa\tnote{2} & 646, 500, 88 & 926 & 500 & 500.3 (0.48) & 500 (500) & 541 (541) & \cite{levy2018innovation} \\ 

& bitcoinAlpha-rev\tnote{3} & 3683, 411, 51 & 22650 & 1738 & 1485.8 (15.9)& 411 (411) & 3251 (3251) & \cite{kumar2016edge} \\
 
 & bitcoinOCT-rev\tnote{3} & 5573, 805, 76 & 32029 & 2883 & 2456 (18.1) & 805 (805) & 5012 (5012) & \cite{kumar2016edge} \\
 
 & Advogato-rev\tnote{4} & 5145, 1148, 729 & 46998 & 1763 & 1759.3 (12.8) & 1148 (1148) & 4362 (4362) & \cite{massa2009bowling} \\
 
& WikiVote-rev\tnote{5} & 10037, 689, 7001 & 139311 & 7097 & 7038.3 (5.7) & 7001 (7001) & 8906 (8906) & \cite{leskovec2010signed} \\

\hline
Airports & US-airport-2010 & 1574, 96, 70 & 28236 & 581 & 556.4 (8.9) & 96 (96) & 1385 (1385) & \cite{opsahl2010node} \\
 & intl.-airport & 2939, 18, 21 & 30501 & 872 & 898.4 (14.0) & 21 (21) & 2547 (2547) & \cite{opsahl2010node} \\ 
\hline
Web & polblogs-rev & 1224, 160, 234 & 19022 & 436 & 354.2 (6.9) & 234 (234) & 992 (992) & \cite{adamic2005political}\\\hline
Power grid & Texas grid & 4889, 379, 1087 & 5855 & 1588 & 1422.4 (12.11) & 1087 (1087) & 2429 (2429) & \cite{bianconi2008local} \\\hline
Neural & C.Elegans & 297, 27, 3 & 2345 & 49 & 28.7 (1.2) & 27 (27) & 179 (179) & \cite{watts1998collective} \\ \hline
p2p & gnutella04 & 10876, 20, 5941 & 39994 & 6004& 5994.8 (6.61) & 5941 (5941) & 7334 (7334) &  \cite{leskovec2007graph} \\
& gnutella05 & 8846, 118, 4996 & 31839 & 5111 & 5032.8 (5.4) & 4996 (4996) & 6074 (6074) &  \cite{leskovec2007graph} \\
& gnutella06 & 8717, 79, 4978 & 31525 & 5033 & 5006.0 (4.56) & 4978 (4978) & 5961 (5961) & \cite{leskovec2007graph}\ \\   \hline    
\end{tabular}

\begin{tablenotes} 

\item[1] This is a fraction of slashdot network named as slashdot-small. We extract  friends relationships and reverse the direction to reflect the influence.

\item[2] The nodes in these networks are growers and advisors in viticulture regions and the link between $i$ and $j$ shows that $i$ communicate with $j$ for viticulture advise. The link directions are reversed.

\item[3] This network is a signed directed network in which the link between $i$ and $j$ shows the rate of trust (-10: total distrust and +10: total trust) of $i$ to $j$. We extract all positive or trust relations and reverse the link direction. 
      
\item[4] The directed link between two developers in the Advogato online community platform of free software, shows a trust. We remove the self-loops and reverse the link directions. 
      
\item[5] Positive votes are extracted and the direction of links are reversed.
       
\end{tablenotes}

\end{threeparttable}
}

\end{table}

\clearpage
\bibliographystyle{IEEEtran} 
\bibliography{refs}
